\documentclass[twocolumn]{aastex631}
\usepackage[utf8]{inputenc}

\usepackage{amsmath}

\usepackage{xcolor}
\usepackage{color}
\definecolor{orange}{rgb}{0.90,0.60,0}
\definecolor{skyblue}{rgb}{0.35,0.70,0.90}
\definecolor{green}{rgb}{0,0.60,0.50}
\definecolor{yellow}{rgb}{0.95,0.90,0.25}
\definecolor{blue}{rgb}{0,0.45,0.70}
\definecolor{vermilion}{rgb}{0.80,0.40,0}
\definecolor{lilac}{rgb}{0.80,0.60,0.70}

\newcommand{\vect}[1]{\boldsymbol{#1}}
\newcommand{\markchange}[1]{ {#1} }

\newcommand{\dif}{\mathrm{d}}
\newcommand{\e}{\mathrm{e}}
\newcommand{\yr}{\,\mathrm{yr}}

\newcommand{\AU}{\,\mathrm{AU}}

\newcommand{\RE}{R_{\oplus}}
\newcommand{\MJ}{M_{\rm J}}
\newcommand{\RJ}{R_{\rm J}}
\newcommand{\MSol}{M_{\odot}}
\newcommand{\RSol}{R_{\odot}}
\newcommand{\LSol}{L_{\odot}}
\newcommand{\kB}{k_{\textsc{b}}}

\newcommand{\ehat}{\vect{\hat{e}}}

\begin{document}

\title{Giant planet engulfment by evolved giant stars: light curves, asteroseismology, and survivability}

\author[0000-0003-3987-3776]{Christopher E.\ O'Connor}
\affiliation{Department of Astronomy and Cornell Center for Astrophysics and Planetary Science, Cornell University, Ithaca, NY 14853, USA}
\affiliation{Kavli Institute for Theoretical Physics, University of California, Santa Barbara, CA 93106, USA}
\correspondingauthor{Chris O'Connor}
\email{coconnor@astro.cornell.edu}

\author[0000-0001-8038-6836]{Lars Bildsten}
\affiliation{Kavli Institute for Theoretical Physics, University of California, Santa Barbara, CA 93106, USA}
\affiliation{Department of Physics, University of California, Santa Barbara, CA 93106, USA}

\author[0000-0002-8171-8596]{Matteo Cantiello}
\affiliation{Center for Computational Astrophysics, Flatiron Institute, New York, NY 10010, USA}
\affiliation{Department of Astrophysical Sciences, Princeton University, Princeton, NJ 08544, USA}

\author[0000-0002-1934-6250]{Dong Lai}
\affiliation{Department of Astronomy and Cornell Center for Astrophysics and Planetary Science, Cornell University, Ithaca, NY 14853, USA}

\shorttitle{Planet engulfment in MESA}
\shortauthors{O'Connor, Bildsten, Cantiello \& Lai}


\received{March 29, 2023}
\revised{May 1, 2023}
\accepted{May 3, 2023}
\submitjournal{AAS Journals}

\begin{abstract}
    About ten percent of Sun-like ($1$--$2 \MSol$) stars will engulf a $1$--$10 \MJ$ planet 
    as they expand during the red giant branch (RGB) or asymptotic giant branch (AGB) phase of their evolution. 
    Once engulfed, these planets experience a strong drag force in the star's convective envelope 
    and spiral inward, depositing energy and angular momentum. 
    For these mass ratios, the inspiral takes $\sim 10$--$10^{2}$ years ($\sim 10^{2}$--$10^{3}$ orbits); 
    the planet undergoes tidal disruption at a radius of $\sim \RSol$. 
    We use the Modules for Experiments in Stellar Astrophysics ({\tt MESA}) software instrument 
    to track the stellar response to the energy deposition while simultaneously evolving the planetary orbit. 
    For RGB stars, as well as AGB stars with $M_{\rm p} \lesssim 5 \MJ$ planets, 
    the star responds quasistatically but still brightens measurably on a timescale of years. 
    In addition, asteroseismic indicators, 
    such as the frequency spacing or rotational splitting, differ before and after engulfment. 
    For AGB stars, engulfment of a $M_{\rm p} \gtrsim 5 \MJ$ planet drives supersonic expansion of the envelope, 
    causing a bright, red, dusty eruption similar to a ``luminous red nova.''
    Based on the peak luminosity, color, duration, and expected rate of these events, 
    we suggest that engulfment events on the AGB could be a 
    significant fraction of low-luminosity red novae in the Galaxy. 
    We do not find conditions where the envelope is ejected prior to the planet's tidal disruption, 
    complicating the interpretation of short-period giant planets orbiting white dwarfs 
    as survivors of common-envelope evolution. 
\end{abstract}

\keywords{Exoplanets (498), 
Red giant stars (1372), 
White dwarf stars (1799), 
Asymptotic giant branch stars (2100), 
Common envelope evolution (2154), 
Star-planet interactions (2177)
}

\section{Introduction} \label{s:Intro}

The engulfment of planets by their host stars occurs a few times per decade in the Galaxy \citep[e.g.][]{Metzger+2012,MacLeod2018}. 
Engulfment can occur during all stages of stellar evolution. 
Around main-sequence (MS) stars, short-period ($P \lesssim 3 \, {\rm days}$) giant planets ($M_{\rm p} \gtrsim 1 \MJ$) 
are vulnerable to tidal destruction on Gyr timescales \citep[e.g.][]{Jackson+2008, Levrard+2009, HS2019}. 
More distant planets can also be ingested after destabilization by a companion 
\citep[e.g.][]{RF1996, Chatterjee+2008, Nagasawa+2008, Naoz+2012, Petrovich2015a, Petrovich2015b, Anderson+2016, Anderson+2020, Stephan+2018}. 
Numerous previous studies have explored 
the potential signatures of planetary engulfment by MS stars, 
including spin-up \citep{Qureshi+2018, Stephan+2020}, 
chemical enrichment \citep{Oh+2018, Spina+2021, Sevilla+2022, Behmard+2022b, Behmard+2022}, 
and both transient and secular brightness changes
\citep{Metzger+2012, Metzger+2017, MacLeod2018}.

Long before the discovery of exoplanets, 
it was recognized that stellar expansion during post-MS evolutionary stages 
would overtake nearby planets \citep[e.g.][]{Alexander1967}. 
The interplay of stellar mass loss, tidal friction, 
and planet--planet interactions during post-MS evolution complicates the question 
of whether and when a given planet will be engulfed \citep{VL2007, VL2009, MV2012, Ronco+2020}. 
Recent discoveries of short-period giant planet candidates orbiting single WDs 
\citep{Vanderburg+2020, Gaia_Arenou+2022} 
\markchange{provide further motivation to investigate the outcomes 
of planetary engulfment by evolved stars} \citep{Lagos+2021, Chamandy+2021, Merlov+2021}.

Based on exoplanet demographics, 
the majority of engulfment events around evolved Sun-like stars involve super-Earths or sub-Neptunes -- 
planets of a predominantly rocky composition, with radii of $1$--$4 \RE$. 
These occur around $\approx 30 \%$ of Sun-like stars on orbital scales of $0.01$--$1 \AU$ \citep[e.g.][]{Fressin+2013, Petigura+2013, Zhu+2018}, 
typically with $3$ to $6$ planets per star 
\citep{Zhu+2018, Zink+2019}. 
They are engulfed during the host star's first ascent of the red giant branch (RGB) 
or on the asymptotic giant branch (AGB) before the onset of thermal pulses. 
However, their small masses have little effect on the host star, 
making their engulfment nearly unobservable 
(apart from the moment of first contact, e.g.\ \citealt{Metzger+2012}).

More impactful engulfment events involve giant planets 
(mass $M_{\rm p} \approx 1$--$10 \MJ$, radius $R_{\rm p} \approx 1 \RJ$). 
Around Sun-like stars, giant planets are relatively scarce at orbital separations less than $1 \AU$, 
with an occurrence fraction of $\approx 3 \%$ \citep{Cumming+2008, Mayor+2011}. 
However, they are more abundant between $1$ and $5 \AU$, 
with an occurrence rate of $\approx 10$--$15\%$ \citep{Fernandes+2019, Fulton+2021}. 
Thus, the majority of the giant planet population around Sun-like stars 
will be engulfed when the host star has a radius of $\gtrsim 1 \AU$ ($= 215 \RSol$): 
near the tip of the RGB or on the AGB \citep{MV2012}. 

\markchange{Many previous studies have explored the effects of an engulfed planet or brown dwarf
on the internal structure and observable properties of a giant star, 
including \citet{LS1984}, \citet{SokerLivioHarpaz1984}, \citet{HS1994}, 
\citet{NelemansTauris1998}, \citet{Soker1998b}, \citet{SL1999, SL1999b}, \citet{RM2003}, 
\citet{Carlberg+2009, Carlberg+2012}, and \citet{Staff+2016}. 
Others have considered the possible role of substellar companions in the origin of hot subdwarf stars \citep[e.g.][]{Soker1998, Nelemans2010, BS2011} 
and single helium-core WDs \citep[e.g.][]{NelemansTauris1998, ZS2022}, 
as well as the morphology of planetary nebulae \citep[e.g.][]{NordhausBlackman2006, Clyne+2014, Boyle2018}. 
Many basic predictions are well established regarding the evolution of the host star during engulfment, 
including large-scale expansion, brightening,
and enhanced mass loss powered by the companion's orbital energy; 
spin-up from the companion's orbital angular momentum; 
and chemical enrichment of the convection zone.}
We revisit the problem, leveraging contemporary improvements 
in input physics and computational methods for 1D stellar models, 
to develop a consistent understanding of planetary engulfment 
and inform the interpretation of recent and future observations.

We focus on the engulfment of so-called `warm Jupiters' and `cold Jupiters', 
giant planets with orbital separations of $0.1$--$1 \AU$ and $>1 \AU$, respectively.
We consider host stars with mass $M_{\star} \approx 1$--$1.5 \MSol$ 
and radius $R_{\star} \approx 50$--$300 \RSol$. 
A parallel study of `hot Jupiter' engulfment by more compact stars 
is being carried out by M.\ Cantiello et al.\ (in prep.).

We conduct numerical simulations of engulfment using the open-knowledge software instrument 
Modules for Experiments in Stellar Astrophysics 
({\tt MESA}: \citealt{Paxton+2011, Paxton+2013, Paxton+2015, Paxton+2018, Paxton+2019, Jermyn+2022}). 
Our main goal is to provide a self-consistent assessment 
of the response of the stellar envelope to energy dissipated by an engulfed planet, 
including potentially observable signatures, for an illustrative range of stellar and planetary properties. 
We also outline the conditions under which a substellar companion can eject the stellar envelope 
and perhaps survive in a short-period post-common-envelope binary with a white dwarf.

In Section \ref{s:Prelim}, we discuss the early stages of planetary engulfment 
and describe our fiducial stellar models. 
In Section \ref{s:InspPhysics}, we summarize the major physical mechanisms 
governing the orbital trajectory of an engulfed planet 
and its eventual tidal disruption deep within the envelope ($r \sim \RSol$). 
In Section \ref{s:StellarResponse}, we discuss the response of the stellar envelope 
to the local deposition of heat by shocks in the vicinity of the inspiraling planet. 
In Section \ref{s:MESA}, we discuss the implementation of planetary engulfment effects in {\tt MESA} and present simulation results. 
In Section \ref{s:Observations}, we evaluate the observability 
of ongoing and previous engulfment events for post-MS stars. 
In Section \ref{s:Conclusion}, we summarize our main findings, discuss open questions, 
and make recommendations for future work.

\section{Preliminaries} \label{s:Prelim}

\subsection{Pre-contact and grazing phases}

Prior to contact between the star and planet, 
two effects bring the planet and the stellar surface closer together: 
stellar radius expansion and orbital decay. 
Orbital decay is driven by tidal friction, most likely due to turbulent dissipation 
in the star's convective envelope \citep[e.g.][]{Zahn1977, Zahn1989, VickLai2020}. 
Drag forces exerted by the stellar corona and wind are negligible \citep{DuncanLissauer1998}. 

When the planet makes contact with the star, drag forces begin to dominate over tidal friction. 
The `grazing' hydrodynamical interaction of the star and planet 
is complex and three-dimensional \citep[cf.][]{ML2020a, Yarza2022}. 
Various observable phenomena, such as expulsion of stellar matter \citep{ML2020, Lau+2022} 
and shock-powered optical and X-ray transients \citep{Metzger+2012, Stephan+2020}, 
may occur during the grazing phase. 
Those are beyond the scope of this study. 
We focus on the later `inspiral' phase of engulfment, 
when the planet is completely immersed in the envelope. 

\subsection{Stellar models}

We use a set of fiducial stellar models obtained with {\tt MESA}-r22.05.1 
(see Section \ref{s:MESA} for software details), 
meant as representative models of RGB and AGB stars. 
We evolved a non-rotating star of initial mass $1.50 \MSol$ 
from the pre-MS through the end of the thermally pulsing AGB stage. 
We adopted an initial chemical composition using the protosolar abundances of \citet{AGS2009}.
We included wind-driven mass loss using the prescriptions 
of \citet{Reimers1975} from the zero-age MS through the RGB 
and \citet{Blocker1995} on the AGB, with respective scaling factors $\eta_{\rm R} = 0.5$ and $\eta_{\rm B} = 0.1$. 
We treated convective boundaries using the Schwarzschild criterion and neglecting overshooting. 
Our stellar models are snapshots of this single evolutionary sequence. 
Table \ref{tab:1.5Msun_models} summarizes their major properties. 
Each model is labeled in the form {\tt ABCxyz}, 
where {\tt ABC} refers to the star's evolutionary stage 
and {\tt xyz} is its approximate radius in units of $\RSol$.

\begin{deluxetable*}{lccccccccccc} \label{tab:1.5Msun_models}
\tablecaption{Properties of fiducial host-star models.}
\tablewidth{0pt}
\tablehead{
\colhead{Model} & \colhead{$R_{\star}$} & \colhead{$M_{\star}$} & \colhead{$M_{\rm core}$} & $E_{\rm grav}$ & $E_{\rm tot}$ & \colhead{$L_{\star}$} & \colhead{$T_{\rm eff}$} & \colhead{$|\lambda|$} & \colhead{$\tau_{\rm dyn}$} & \colhead{$\tau_{\rm KH}$} & \colhead{$\mathcal{N}(3 \MJ)$} \\
        & \colhead{[$\RSol$]} & \colhead{[$\MSol$]} & \colhead{[$\MSol$]} & \colhead{[erg]} & \colhead{[erg]} & \colhead{[$\LSol$]} & \colhead{[K]} & \colhead{--} & \colhead{[yr]} & \colhead{[yr]} & \colhead{--}
}

\startdata
        {\tt RGB50} & 50.1 & 1.493 & 0.362 & $-2.1 \times 10^{47}$ & $-6.9 \times 10^{46}$ & 549 & 3950 & 1.8 & 0.015 & 2500 & 150 \\ 
        {\tt RGB100} & 100.0 & 1.420 & 0.424 & $-1.1 \times 10^{47}$ & $-2.4 \times 10^{46}$ & 1511 & 3599 & 2.2 & 0.042 & 420 & 340 \\ 
        {\tt RGB150} & 149.3 & 1.357 & 0.474 & $-7.1 \times 10^{46}$ & $-9.9 \times 10^{45}$ & 2672 & 3396 & 3.1 & 0.079 & 150 & 550 \\ 
        {\tt AGB200} & 203.7 & 1.278 & 0.555 & $-4.7 \times 10^{46}$ & $-3.5 \times 10^{45}$ & 4124 & 3242 & 4.9 & 0.13 & 60 & 833 \\ 
        {\tt AGB275} & 276.7 & 1.000 & 0.568 & $-1.8 \times 10^{46}$ & $+1.5 \times 10^{45}$ & 5602 & 3003 & 4.0 & 0.23 & 20 & 1950 \\ 
\enddata

\tablecomments{\markchange{All stellar models were evolved from a $1.50 \MSol$ ZAMS model.}
Definitions: $M_{\rm core}$ is the enclosed mass at the bottom of the convection zone. 
The global dynamical time $\tau_{\rm dyn} \equiv (R_{\star}^{3} / G M_{\star})^{1/2}$ 
and Kelvin--Helmholtz time $\tau_{\rm KH} \equiv G M_{\star}^{2} / ( R_{\star} L_{\star} )$. 
The gravitational binding energy of the envelope is $E_{\rm grav}$. 
The total binding energy $E_{\rm tot}$ and factor $\lambda$ are defined in Eq.\ (\ref{eq:def_Ebind}). 
The total number of orbits undergone by a $3 \MJ$ giant planet is $\mathcal{N}(3 \MJ)$.}

\end{deluxetable*}

\section{Physics of inspiral} \label{s:InspPhysics}

In this section, we summarize the local hydrodynamical interaction 
between an engulfed giant planet and the envelope of a typical late-stage giant star, 
with an emphasis on the drag force exerted on the planet and the associated energy deposition. 
We describe the planet's trajectory in a `passive' stellar envelope 
and its eventual tidal disruption.

\subsection{Drag force} \label{s:InspPhysics:Drag}

\begin{figure}
    \centering
    \includegraphics[width=\columnwidth]{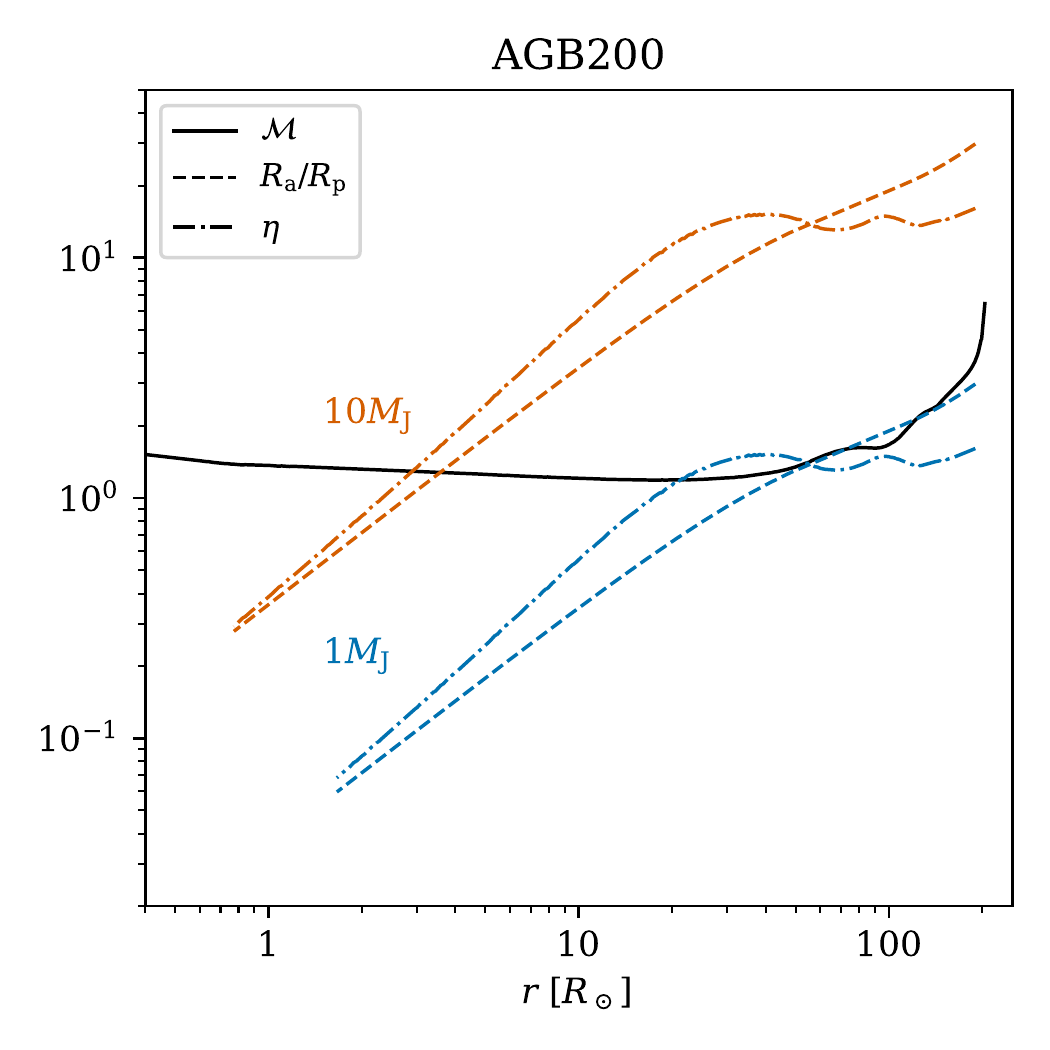}
    \caption{Profiles of the quantities $\mathcal{M}$ (solid black curve; Eq.\ \ref{eq:def_Mach}), 
    $R_{\rm a}/R_{\rm p}$ (dashed curves; Eq.\ \ref{eq:def_Ra}),
    and $\eta$ (dot-dashed curves; Eq.\ \ref{eq:def_eta_hydro}) 
    within stellar model {\tt AGB200}. 
    Blue and vermilion curves are for $1 \MJ$ and $10 \MJ$ planets, respectively, with $R_{\rm p} = 1 \RJ$. 
    Dashed and dot-dashed curves are terminated at the planet's Roche limit (Eq.\ \ref{eq:def_adis}). 
    The bottom of the stellar convection zone is at $r \approx 0.8 \RSol$.}
    \label{fig:MachRshock_profiles}
\end{figure}

The drag force acting on an engulfed planet is determined 
by the flow of stellar matter around it. 
We assume that the planet instantaneously follows a circular orbit 
with radius $a$ and velocity $\vect{v} = v_{\rm K}(a) \ehat_{\phi}$, with
\begin{equation} \label{eq:vKepler}
    v_{\rm K} = \left( \frac{G M_{r}}{r} \right)^{1/2},
\end{equation}
where $M_{r}$ is the stellar mass enclosed by a shell of radius $r$. 
The planet's Mach number is
\begin{equation} \label{eq:def_Mach}
    \mathcal{M} = v_{\rm K}/c_{s}, 
\end{equation}
where $c_{s}$ is the local speed of sound in the stellar envelope. 
We can safely assume that $v_{\rm K}$ and $c_{s}$ are much greater 
than the bulk velocity of fluid near the planet due to stellar rotation and convection. 
The planet's motion is always supersonic, with $1 \lesssim \mathcal{M} \lesssim 10$ in general 
and $1 \lesssim \mathcal{M} \lesssim 2$ in the deepest portion of the envelope 
(where the most energy is dissipated). 
The run of $\mathcal{M}$ for stellar model {\tt AGB200}
is shown by the solid curve in Figure \ref{fig:MachRshock_profiles}.

Despite the supersonic motion through the envelope, 
the interior of the inspiraling giant planet remains in hydrostatic equilibrium for most of the inspiral
because the ram pressure exerted by the medium is small compared to the planet's central pressure. 
The drag on the planet comprises 
hydrodynamic friction (a.k.a.\ ram pressure or geometric friction), 
which arises from the non-uniform fluid stress exerted across the object's surface; 
and gravitational dynamical friction, 
which arises from the wake formed behind the object by gravitational focusing. 
The importance of each component is determined by the planet's radius $R_{\rm p}$ 
and gravitational focusing (a.k.a.\ Bondi--Hoyle) radius
\begin{equation} \label{eq:def_Ra}
    R_{\rm a} = \frac{2 G M_{\rm p}}{v^{2}},
\end{equation}
where $M_{\rm p}$ is the planet's mass.  
For an object with velocity $\vect{v}$ moving through a uniform medium of density $\rho$, 
the hydrodynamic and gravitational drag forces may be written
\begin{align}
    \vect{F}_{\rm d} &= - C_{\rm d} \pi R_{\rm p}^{2} \rho v \vect{v} \propto v^{2}, \label{eq:def_Fdrag} \\
    \vect{F}_{\rm g} &= - C_{\rm g} \pi R_{\rm a}^{2} \rho v \vect{v} \propto \frac{1}{v^{2}}. \label{eq:def_Fgrav}
\end{align}
The coefficients $C_{\rm d}$ and $C_{\rm g}$ (see below) are determined by the global flow structure, 
which depend on the dimensionless quantities $\mathcal{M}$ and $R_{\rm a}/R_{\rm p}$. 
Note that our definition of $C_{\rm d}$ absorbs 
a leading factor $(1/2)$ that others prefer to be explicit. 
All else being equal, gravitational (hydrodynamical) drag dominates 
when $v$ is small (large), which tends to occur 
at large (small) $r$ within the envelope. 
The transition between the drag regimes occurs 
when $v$ is of the order of the planet's surface escape velocity, 
$v_{\rm e} = (2 G M_{\rm p} / R_{\rm p})^{1/2}$.

\citet{Thun2016} and \citet{Yarza2022} studied the flow 
around a gravitating supersonic projectile 
using 2D-axisymmetric and 3D hydrodynamics simulations, respectively. 
Both used a `wind tunnel' setup to characterize the steady-state flow structure 
and numerically derive the hydrodynamic and gravitational drag forces 
on the projectile (see also \citealt{MacLeod2017}). 
For supersonic motion, a shock develops ahead of the projectile. 
The structure of the post-shock material depends on the Mach number 
and the `compactness' parameter $R_{\rm a} / R_{\rm p}$. 
\citet{Thun2016} found that the stand-off distance $R_{\rm sh}$ of the shock 
from the center of the projectile is given approximately by
\begin{equation} \label{eq:def_Rshock}
    R_{\rm sh} = R_{\rm p} \max\left( 1, \eta \right),
\end{equation}
where
\begin{equation} \label{eq:def_eta_hydro}
    \eta \equiv \frac{g(\gamma)}{2} \frac{\mathcal{M}^{2}}{\mathcal{M}^{2} - 1} \frac{R_{\rm a}}{R_{\rm p}}
\end{equation} 
is a nonlinearity parameter \citep{KK2009}
and the factor $g(\gamma)$ depends on the adiabatic exponent $\gamma$. 
We take $\gamma = 5/3$ and $g(5/3)=1$ throughout this work.
The hydrodynamic crossing time ($\sim R_{\rm sh} / v$) 
is short compared to the Keplerian orbital period and the orbital decay time, 
so we assume that the shock structure is in a steady state determined by local conditions. 
Fig.\ \ref{fig:MachRshock_profiles} shows the run of $R_{\rm a}/ R_{\rm p}$ and $\eta$ for model ${\tt AGB200}$ 
for two planet masses, $M_{\rm p} = 1 \MJ$ and $10 \MJ$.

When gravitational focusing is negligible ($R_{\rm a} \ll R_{\rm p}$), 
the post-shock material is compressed ahead of the projectile and rarefied behind it. 
The total drag force is then dominated by hydrodynamic friction. 
The drag coefficient is weak function of Mach number 
and can be estimated using laboratory ballistics data \citep{BH1972}. 
We use the following fitting formula:
\begin{equation} \label{eq:Cdrag_approx}
    C_{\rm d} = 0.375 + 0.125 \tanh[ 1.75 (\mathcal{M} - 1) ],
\end{equation}
which provides an adequate estimate for a non-gravitating projectile 
with Reynolds number $\gg 1$ (appropriate for our case).

In the limit of strong gravitational focusing ($R_{\rm a} \gg R_{\rm p}$), 
post-shock material accumulates around the projectile, 
creating a quasi-spherical `halo' of subsonic matter 
with radius $\approx R_{\rm a}$ \citep{Thun2016}. 
This structure arises from the impenetrable boundary condition at the projectile's surface; 
it does not occur in simulations with absorbing 
or outflowing boundary conditions \citep[e.g.][]{MacLeod2017, XLi2020}. 
The halo is approximately in hydrostatic equilibrium 
and exerts almost uniform pressure on the projectile; 
thus hydrodynamic drag is negligible ($C_{\rm d} \simeq 0$) in this limit. 
The gravitational drag coefficient is given by \citep{Ostriker1999, Thun2016}:
\begin{equation} \label{eq:Cgrav}
    C_{\rm g} = \ln\left( \frac{R_{\rm max}}{R_{\rm sh}} \right) - \frac{1}{2} \ln\left( \frac{\mathcal{M}^{2}}{\mathcal{M}^{2} - 1} \right),
\end{equation}
where $R_{\rm max}$ is the maximum extent of the wake formed behind the sphere.

In order for the `wind tunnel' results of \citet{Thun2016} to be applicable 
to the motion of a body orbiting within a stratified stellar envelope, 
$R_{\rm max}$ must not exceed the local density scale height, $H_{\rho}$. 
The importance of the density gradient is measured by the parameter
\begin{equation} \label{eq:def_epsRho}
    \epsilon_{\rho} = \frac{R_{\rm sh}}{H_{\rho}}.
\end{equation}
\citet{Yarza2022} conducted wind-tunnel simulations with $0.1 \leq \epsilon_{\rho} \leq 1$ 
and $0.3 \leq R_{\rm p} / R_{\rm a} \leq 1$. 
They found that the presence of a density gradient 
introduces asymmetry to the flow structure around the object 
but otherwise does not change the main conclusions of \citet{Thun2016}. 
Notably, the total drag force increases by less than a factor of $\sim 2$ 
relative to that exerted by a uniform medium. 
For our cases, we find that $\epsilon_{\rho} \lesssim 0.2$ for $r \lesssim 0.95 R_{\star}$, 
indicating that corrections to the local drag-force law 
due to the envelope's density gradient are negligible for giant planets whose `spheres of influence' 
($\sim$ few $R_{\rm sh}$) are well separated from the stellar surface.

The results summarized above indicate that, 
because the structure of the flow around the planet depends mainly on $R_{\rm a}/R_{\rm p}$, 
the hydrodynamic and gravitational regimes of the drag force are almost mutually exclusive.
We adopt the following prescription for the total drag force during the inspiral phase:
\begin{equation} \label{eq:Fdrag_for_MESA}
    \vect{F}_{\rm drag} = - \ehat_{\phi} \max\left[ C_{\rm d} \pi R_{\rm p}^{2}, C_{\rm g} \pi R_{\rm a}^{2} \right] \rho v_{\rm K}^{2},
\end{equation}
where $C_{\rm d}$ is given by Eq.\ (\ref{eq:Cdrag_approx}) 
and $C_{\rm g}$ by Eq.\ (\ref{eq:Cgrav}) with $R_{\rm max} = H_{\rho}$. 

\subsection{Orbital decay and heat deposition} \label{s:InspPhysics:OrbDecayHeat}

The rate of orbital decay is directly related to the rate of non-conservative work by drag forces:
\begin{equation} \label{eq:adot}
    \frac{\dot{a}}{a} = \frac{\vect{F}_{\rm drag} \cdot \vect{v}}{|E_{\rm orb}|} \equiv - \frac{1}{\tau_{\rm insp}},
\end{equation}
where
\begin{equation} \label{eq:def_Eorb}
    E_{\rm orb} = - \frac{G M_{r}(a) M_{\rm p}}{2 a}
\end{equation}
and $\tau_{\rm insp}$ is a characteristic inspiral time at a given separation. 
We assume that the orbital energy loss is deposited in the vicinity of the planet as heat. 
The heat deposition rate is
\begin{equation} \label{eq:def_Ldrag}
    L_{\rm drag} = - \vect{F}_{\rm drag} \cdot \vect{v} = - \dot{E}_{\rm orb}.
\end{equation}
Heat is deposited at the shock that develops around the planet \citep{Thun2016, Yarza2022} 
and diffuses \markchange{via convection into the surroundings \citep[e.g.][]{WilsonNordhaus2022}.} 
For further discussion of the role of stellar convection, see Section \ref{s:StellarResponse}.

When gravitational dynamical friction dominates, the inspiral timescale is roughly
\begin{align} \label{eq:tinsp_grav}
    \tau_{\rm insp} \sim \frac{1}{C_{\rm g}} \frac{M_{r}}{M_{\rm p}} \tau_{\rm orb},
\end{align}
where $\tau_{\rm orb} = 2 \pi (a^{3} / G M_{r})^{1/3}$ is the local Keplerian orbital period. 
When hydrodynamic drag dominates, it is roughly
\begin{align} \label{eq:tinsp_drag}
    \tau_{\rm insp} \sim \frac{1}{C_{\rm d}} \frac{M_{\rm p}}{M_{r}} \left( \frac{a}{R_{\rm p}} \right)^{2} \tau_{\rm orb}.
\end{align}
In all cases we consider, the planet undergoes $\mathcal{N} \sim 10^{2}$--$10^{3}$ orbits in total during the inspiral. 

\subsubsection{Angular momentum deposition} \label{s:InspPhysics:OrbDecayHeat:SpinUp}

Drag forces also exert a torque that reduces the planet's orbital angular momentum 
and spins up the star. 
In principle, one could study the differential rotation 
induced by deposition of the planet's angular momentum 
in successive layers of the star using {\tt MESA}. 
Since our focus is the effect of energy deposition by the planet, 
we neglect spin-up in our calculations. 
It suffices to say that the engulfment would greatly increase the star's rotation rate 
\citep[e.g.][]{LS2002, Carlberg+2009, Carlberg+2012, Privitera+2016, Stephan+2020}. 
Assuming the star spins up rigidly and has negligible initial angular momentum, 
the post-engulfment rotation rate is
\begin{equation}
    \Omega_{\star} = \frac{1}{k_{\star}} \frac{M_{\rm p}}{M_{\star}} \Omega_{\rm K\star},
\end{equation}
where $\Omega_{\rm K\star}$ is the Keplerian angular velocity at the stellar surface (a.k.a.\ the breakup rate) 
and where the stellar moment of inertia is $k_{\star} M_{\star} R_{\star}^{2}$ with $k_{\star} \approx 0.1$. 
For $M_{\rm p} = 1$--$10 \MJ$ and $M_{\star} \approx \MSol$, 
final rotation rates $\approx 0.01$--$0.1 \Omega_{\rm K\star}$ are expected. 
Such rapid rotation would be a lasting signature of engulfment 
detectable via spectroscopic broadening, photometric modulation, and asteroseismic mode splitting. 
\markchange{It could also generate magnetic dynamo activity in the envelope \citep[e.g.][]{LS2002, NordhausBlackman2006}.}

\subsection{Disruption of the planet} \label{s:InspPhysics:Disrupt}

Engulfed planets are usually destroyed in the stellar interior. 
This can occur either catastrophically or gradually, 
depending on the planet's properties and the envelope conditions. 

\citet{LS1984} proposed that an engulfed substellar body is thermally disrupted (``dissolved'') 
upon reaching a location where the local sound speed exceeds the body's surface escape velocity. 
Though frequently quoted in the literature  
\citep[e.g.][]{SL1999, SL1999b, Carlberg+2009, Aguilera+2016, Privitera+2016, Lau+2022, Cabezon+2022}, 
this criterion is only the first step needed for dissolution. 
The second is to compare the inspiral time with the timescale 
for heating of the planetary interior by the ambient medium. 
For irradiated giant planets, the rate-limiting processes are 
conductive and radiative heat transfer, 
as the rising entropy of the outer layers suppresses convection \citep{Guillot+1996, AB2006}. 
This implies a heating time far exceeding the predicted inspiral time; 
hence, we do not consider thermal disruption to be relevant for giant planets. 

Alternatively, planets can be disrupted by ram pressure or ablation 
due to their supersonic motion in a stellar envelope \citep{JS2018}. 
Neither of these is relevant for us because the planet 
is much denser than its surroundings, even near the base of the convection zone. 

Based on these considerations, we assume that an engulfed planet survives 
in the stellar envelope until it fills its Roche lobe, 
whereupon it is undergoes rapid tidal disruption 
\citep[cf.][]{ReyesRuizLopez1999, NordhausBlackman2006, Nordhaus+2011, Guidarelli+2022}. 
The orbital radius $a_{\rm dis}$ where this occurs 
is \citep{Eggleton1983}
\begin{align} \label{eq:def_adis}
    a_{\rm dis} &\simeq 2 R_{\rm p} \left( \frac{M_{\rm core}}{M_{\rm p}} \right)^{1/3} \\
    &\approx 1.8 \RSol \left( \frac{R_{\rm p}}{\RJ} \right) \left( \frac{M_{\rm core}}{0.5 \MSol} \frac{\MJ}{M_{\rm p}} \right)^{1/3}, \nonumber
\end{align}
where $M_{\rm core}$ is the mass of the compact stellar core. 
Planets are disrupted long before making physical contact with the core ($r \lesssim 0.1 \RSol$). 
We do not consider the subsequent evolution of tidal debris in this work. 
However, previous studies have considered various possibilities 
\citep{ReyesRuizLopez1999, SL1999, SL1999b, NordhausBlackman2006, Nordhaus+2011, Guidarelli+2022}.

\subsection{Energy budget considerations} \label{s:InspPhysics:Energetics}

When a planet undergoes inspiral from an initial separation $a_{1} \simeq R_{\star}$ to a final separation $a_{2} \ll a_{1}$, 
the total energy deposited in the envelope is given by
\begin{equation} \label{eq:def_W}
    W = \left| - \frac{G M_{r}(a_{2}) M_{\rm p}}{2 a_{2}} + \frac{G M_{\star} M_{\rm p}}{2 a_{1}} \right | \simeq \frac{G M_{\rm core} M_{\rm p}}{2 a_{2}},
\end{equation}
setting the overall energy budget for the inspiral. 
If heat deposition ceases (or at least slows down) when the planet is tidally disrupted, 
then by setting $a_{2} = a_{\rm dis}$, we find an estimate of the characteristic energy budget:
\begin{align}
    W_{\rm insp} &= \frac{G (M_{\rm core}^{2} M_{\rm p}^{4})^{1/3}}{4 R_{\rm p}} \nonumber \\
    &\approx 5.6 \times 10^{44} \, {\rm erg} \nonumber \\
    & \hspace{0.5cm} \times \left( \frac{R_{\rm p}}{\RJ} \right)^{-1} \left( \frac{M_{\rm core}}{0.5 \MSol} \right)^{2/3} \left( \frac{M_{\rm p}}{\MJ} \right)^{4/3}. \label{eq:W_budget}
\end{align}
This should be compared with the total binding energy of the stellar envelope:
\begin{equation} \label{eq:def_Ebind}
    E_{\rm tot} = \int_{M_{\rm core}}^{M_{\star}} \left( - \frac{G M_{r}}{r} + u \right) \dif M_{r} \equiv - \frac{G M_{\star} M_{\rm env}}{\lambda R_{\star}},
\end{equation}
where $M_{\rm env} = M_{\star} - M_{\rm core}$ is the mass of the envelope, 
and $u$ is the specific internal energy of stellar gas. 
Although the disruption radius $a_{\rm dis}$ varies as a function of planet mass, 
the mass enclosed by the terminal orbit is always very close to $M_{\rm core}$. 
Table \ref{tab:1.5Msun_models} lists the values of $E_{\rm tot}$ and $\lambda$ for our fiducial stellar models.
We also give the gravitational binding energy $E_{\rm grav}$ of the envelope,
computed by ommiting $u$ in Eq.\ (\ref{eq:def_Ebind}).

Both $|E_{\rm tot}|$ and $|E_{\rm grav}|$ range from $10^{45}$ to $10^{47} \, {\rm erg}$ in order of magnitude. 
The envelope's $E_{\rm tot}$ is usually negative ($\lambda > 0$) 
but becomes positive ($\lambda < 0$) on the upper AGB, 
where ionization energy overwhelms gravitational potential energy \citep{PZ1968}.

Comparison of $W_{\rm insp}$ and $|E_{\rm tot}|$ gives a sense 
of whether the deposition of heat by the planet leads to significant changes 
in the overall structure of the stellar envelope during inspiral. 
By setting $W_{\rm insp} = |E_{\rm tot}|$ and solving for $M_{\rm p}$, 
we obtain a characteristic planetary mass for which we expect a major structural adjustment:
\begin{align}
    M_{\rm p,crit} &= \left( \frac{4}{|\lambda|} \frac{R_{\rm p}}{R_{\star}} \right)^{3/4} \frac{(M_{\star} M_{\rm env})^{3/4}}{M_{\rm core}^{1/2}} \nonumber \\
    &\approx 2.8 \MJ \left( \frac{4}{|\lambda|} \frac{R_{\rm p}}{\RJ} \frac{200 \RSol}{R_{\star}} \right)^{3/4} \nonumber \\
    & \hspace{0.5cm} \times \left( \frac{M_{\star}}{\MSol} \frac{M_{\rm env}}{0.5 \MSol} \right)^{3/4} \left( \frac{0.5 \MSol}{M_{\rm core}} \right)^{1/2}. \label{eq:def_Mpcrit}
\end{align}
However, as we show in Section \ref{s:MESA}, 
energetic arguments alone cannot predict the range of possible outcomes. 

\markchange{Our calculation of the inspiral energy budget is somewhat incomplete 
in that we ignore the evolution of the debris from the planet's tidal disruption (see previous section). 
The continued release of gravitational potential energy by the sinking debris \citep[e.g.][]{SL1999, SL1999b} 
would contribute additional energy $W_{\rm sink} \sim W_{\rm insp}$ 
over a timescale $\tau_{\rm sink}$, 
assuming that the debris sinks to a final radius $\sim r_{\rm tide}/2$. 
Since the central temperature of a gas giant or brown dwarf 
is typically much less than the ambient temperature $\sim T(r_{\rm tide})$, 
the debris would also absorb energy $(-W_{\rm abs})$ as heat 
from the environment on a timescale $\tau_{\rm abs}$ \citep{HS1994}. 
The quantity of heat absorbed is
\begin{align}
    |W_{\rm abs}| &\sim \frac{M_{\rm p}}{m_{p}} k_{\rm B} T(r_{\rm tide}) \nonumber \\
    &\approx 2 \times 10^{44} \, {\rm erg} \left( \frac{M_{\rm p}}{\MJ} \right) \left( \frac{T(r_{\rm tide})}{10^{6} \, {\rm K}} \right),
\end{align}
where $m_{p}$ is the mass of a proton and $k_{\rm B}$ is Boltzmann's constant. 
This happens also to be of the order of $W_{\rm insp}$ in the cases we consider. 
These additional process act as residual sources and sinks of energy 
after a planetary inspiral. 
The timescales $\tau_{\rm sink}$ and $\tau_{\rm abs}$ are somewhat uncertain; 
they depend in detail on the dynamical evolution of the tidal debris. 
If either timescale is comparable to, or shorter than, the inspiral time 
just before tidal disruption ($\tau_{\rm late}$; see Section \ref{s:InspPhysics:LateInsp}), 
these effects could potentially alter the evolution of the host star to a significant degree. 
On the other hand, if $\tau_{\rm abs} \approx \tau_{\rm sink}$, 
then the effects may cancel one another because $W_{\rm insp} \sim -W_{\rm orb}$.
It would be possible to incorporate these effects in {\tt MESA} 
using a parametrized prescription for the sinking and thermal evolution of tidal debris. 
We leave this for future work.}

\subsection{The passive-envelope approximation} \label{s:InspPhysics:PassEnv}

A crude but useful approximation of the inspiral trajectory 
can be obtained by assuming that an engulfed planet has no effect on the stellar envelope. 
We call this the `passive-envelope' approximation. 
As the inspiral is brief compared to the stellar evolutionary timescale, 
a snapshot of the stellar structure is sufficient to compute the planet's trajectory.

The right-hand side of Eq.\ (\ref{eq:adot}) determines the inspiral rate $\dot{a}$ 
in terms of the properties of the envelope and planet. 
In the passive-envelope approximation, the planet's orbital separation $a$ at time $t$ is given by
\begin{equation} \label{eq:traj_PassEnv}
    t - t_{1} = - \int_{a}^{a_{1}} \frac{\dif r}{\dot{a}(r; M_{\rm p}, R_{\rm p})},
\end{equation}
where $a_{1}$ is the separation of the planet at time $t_{1}$. 
The instantaneous heating rate due to drag ($L_{\rm drag}$) 
along the planet's trajectory can also be computed.

\subsection{The importance of the `late' inspiral} \label{s:InspPhysics:LateInsp}

\begin{figure}
    \centering
    \includegraphics[width=\columnwidth]{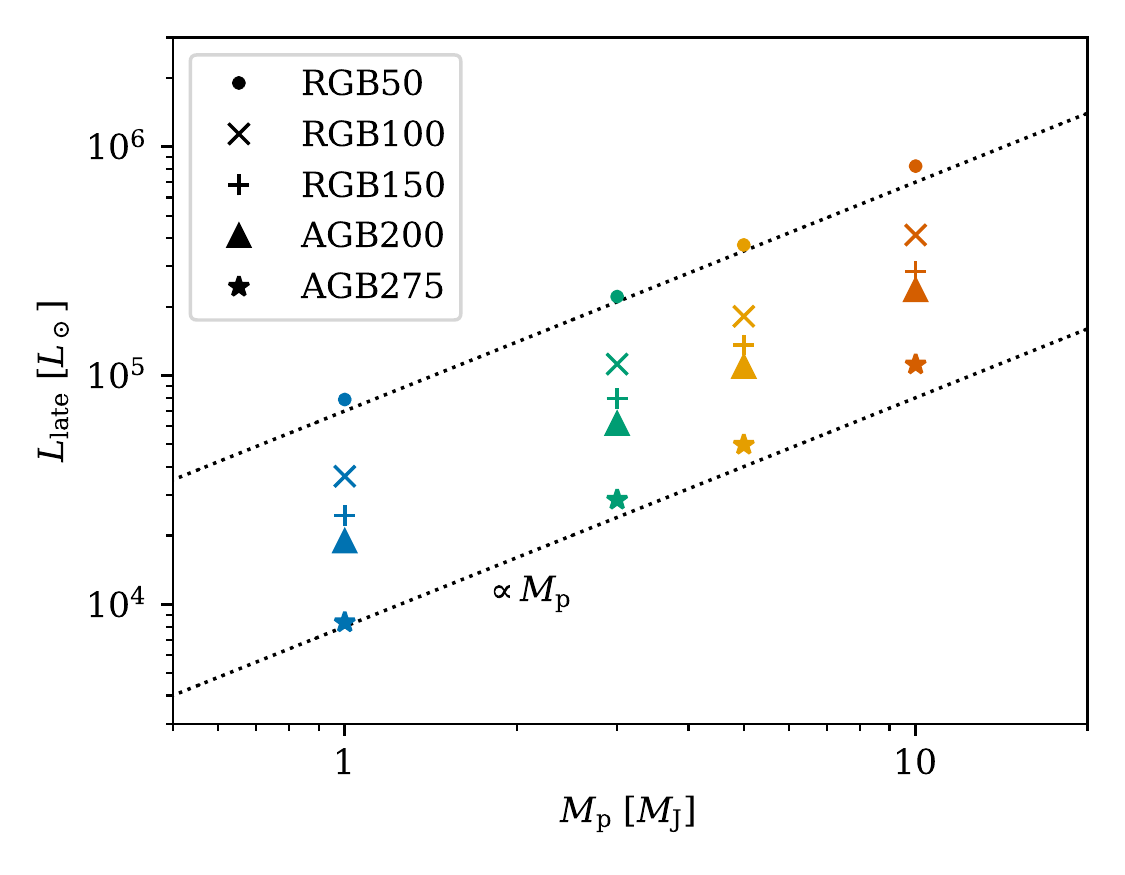}
    \caption{Heating rate during the late inspiral 
    as a function of planetary mass for different stellar models 
    in the passive-envelope approximation. 
    The dotted lines show a linear scaling with $M_{\rm p}$ for comparison.}
    \label{fig:Llate_passive}
\end{figure}

As the planet spirals inward, the cumulative heat deposited in the star grows like $1/a$. 
Roughly half of the total heat is deposited between $r = 2 a_{\rm dis}$ and $r = a_{\rm dis}$. 
We dub this the `late' stage of inspiral. 
This stage is critical for determining the qualitative behavior of the stellar envelope response. 
We denote the energy deposited in the envelope during this stage as $W_{\rm late} \simeq W_{\rm max}/2$. 
The duration of the late inspiral, denoted $\tau_{\rm late}$, 
is of the order of the local inspiral time $\tau_{\rm insp}$ evaluated at $r = 2 a_{\rm dis}$; 
in the passive-envelope approximation, $\tau_{\rm late}$ 
can be computed exactly by way of Eq.\ (\ref{eq:traj_PassEnv}). 
The average heating rate due to drag during the late inspiral 
is $L_{\rm late} \equiv W_{\rm late} / \tau_{\rm late}$.

In Figure \ref{fig:Llate_passive}, we show $L_{\rm late}$ as a function of $M_{\rm p}$ 
for all stellar models, as computed under the passive-envelope approximation. 
For a given star, $L_{\rm late}$ often exceeds the intrinsic luminosity $L_{\star}$ 
and is nearly proportional to $M_{\rm p}$. 
The latter fact implies that $\tau_{\rm late}$ depends weakly on $M_{\rm p}$, 
in qualitative agreement with the scaling $\tau_{\rm insp} \propto M_{\rm p} a_{\rm dis}^{2} \propto M_{\rm p}^{1/3}$ 
predicted by Eqs.\ (\ref{eq:tinsp_drag}) and (\ref{eq:def_adis}). 
Our {\tt MESA} experiments confirm that the passive-envelope approximation 
predicts the planet's trajectory accurately in many cases; 
they also allow for a prediction of when the approximation breaks down.

\section{Stellar response} \label{s:StellarResponse}

The planet's supersonic motion through the stellar envelope 
dissipates orbital energy (via shocks) as heat along its trajectory. 
The planet acts like an embedded heat source, 
localized in a region of characteristic size $R_{\rm sh}$ (Eq.\ \ref{eq:def_Rshock}), 
emitting an instantaneous power $L_{\rm drag}$ determined by local properties of the envelope. 
The structural response of the stellar envelope to this energy deposition 
may be determined in a first approximation 
via a local analysis of heat transport in the planet's vicinity.

The dominant mode of heat transport in the envelope of a giant star is convection, 
which is treated using the traditional mixing-length theory (MLT). 
Turbulence causes heat diffusion on a local eddy turnover timescale
\begin{equation} \label{eq:def_teddy}
    \tau_{\rm eddy} = \frac{\alpha_{\rm MLT} H_{P}}{v_{\rm c}},
\end{equation}
where $H_{P} (\sim H_{\rho})$ is the local pressure scale height, 
$v_{\rm c}$ is the convective turbulent velocity, and $\alpha_{\rm MLT}$ is a free parameter. 
In this section, we set $\alpha_{\rm MLT} = 1$ for convenience; 
we use $\alpha_{\rm MLT} = 2$ in our {\tt MESA} simulations. 
The quantity $\tau_{\rm eddy}$ will be compared with local dynamical timescales, namely the sound propagation time
\begin{equation} \label{def_tsound}
    \tau_{\rm sound} = \frac{H_{P}}{c_{s}}
\end{equation}
and the Keplerian orbital period $\tau_{\rm orb}$. 
In most of the envelope, these timescales have a hierarchy 
$\tau_{\rm sound} \ll \tau_{\rm orb} \lesssim \tau_{\rm eddy}$. 
\markchange{
All of these timescales are short compared to the inspiral time $\tau_{\rm insp}$. 
For further discussion of the role of convection in common-envelope events 
(and planetary engulfment events), 
we refer the reader to some recent works by \citet{Sabach+2017}, \citet{Grichener+2018}, 
and \citet{WilsonNordhaus2019, WilsonNordhaus2020, WilsonNordhaus2022}.
}

We can reasonably describe heat transport near the planet as follows:  
In the immediate wake of the companion, fluid elements compressed and heated by the shock 
re-expand on a timescale $\tau_{\rm sound}$. 
After reaching a size of order $H_{P}$, the heated region equilibrates 
and loses excess heat to its surroundings. 
This implies that, over the orbital timescale, 
the companion directly heats a toroidal region of major radius $a$ and minor radius $H_{P}$. 
However, over the longer timescale $\tau_{\rm eddy}$, fluid motions 
due to both convection and the repeated passages of the companion 
should redistribute the excess heat roughly along isobars.\footnote{
The actual rate of lateral heat transport in a convective layer is uncertain. 
If the corresponding velocity is not too different from $v_{\rm c}$, 
the lateral transport time $\sim r/v_{\rm c}$ is only a factor of a few longer than $\tau_{\rm eddy}$. 
This is because $H_{P}$ is only a factor of few smaller than $r$ 
deep inside an RGB or AGB envelope.} 
{\it In other words, over timescales longer than $\tau_{\rm eddy}$, 
we may consider the heat dissipated by drag to be deposited 
inside a spherical shell of radius $a$ and thickness $H_{P}$.} 
This is our `shellular heating' approximation.

Assuming uniform heat deposition per unit mass in the heated shell, 
the heating rate per unit mass is
\begin{equation} \label{eq:def_eps_drag}
    \varepsilon_{\rm drag} = \frac{L_{\rm drag}}{\Delta M_{r}} \simeq \frac{L_{\rm drag}}{4 \pi a^{2} H_{P}(a) \rho(a)}, \ \ a - \frac{H_{P}}{2} < r < a + \frac{H_{P}}{2},
\end{equation}
and zero elsewhere. 
We define a local heating timescale
\begin{equation} \label{eq:def_theat}
    \tau_{\rm heat} = \frac{c_{P} T}{\varepsilon_{\rm drag}},
\end{equation}
where $c_{P}$ is the envelope's 
specific heat at constant pressure and $T$ its ambient temperature. 
We also define a local thermal cooling time for comparison:
\begin{equation} \label{eq:def_tcool}
    \tau_{\rm therm} = \frac{4 \pi r^{2} H_{P} \rho c_{P} T}{L_{\star}} \simeq \tau_{\rm heat} \left( \frac{L_{\rm drag}}{L_{\star}} \right),
\end{equation}
where $L_{\star}$ is the star's intrinsic luminosity. 
As we will show, when $L_{\rm drag} \ll L_{\star}$, the shell heated by the planet 
remains nearly in thermal equilibrium with the rest of the stellar envelope; 
the star can accommodate the additional luminosity with a negligible adjustment of its thermal structure. 
Conversely, when $L_{\rm drag} \gtrsim L_{\star}$, 
the heated shell is no longer in thermal equilibrium 
and expands on a timescale $\tau_{\rm heat}$. 
Provided that $\tau_{\rm sound} \ll \tau_{\rm heat}$, 
thermal expansion is locally hydrostatic and occurs nearly at constant pressure.

\begin{figure}
    \centering
    \includegraphics[width=\columnwidth]{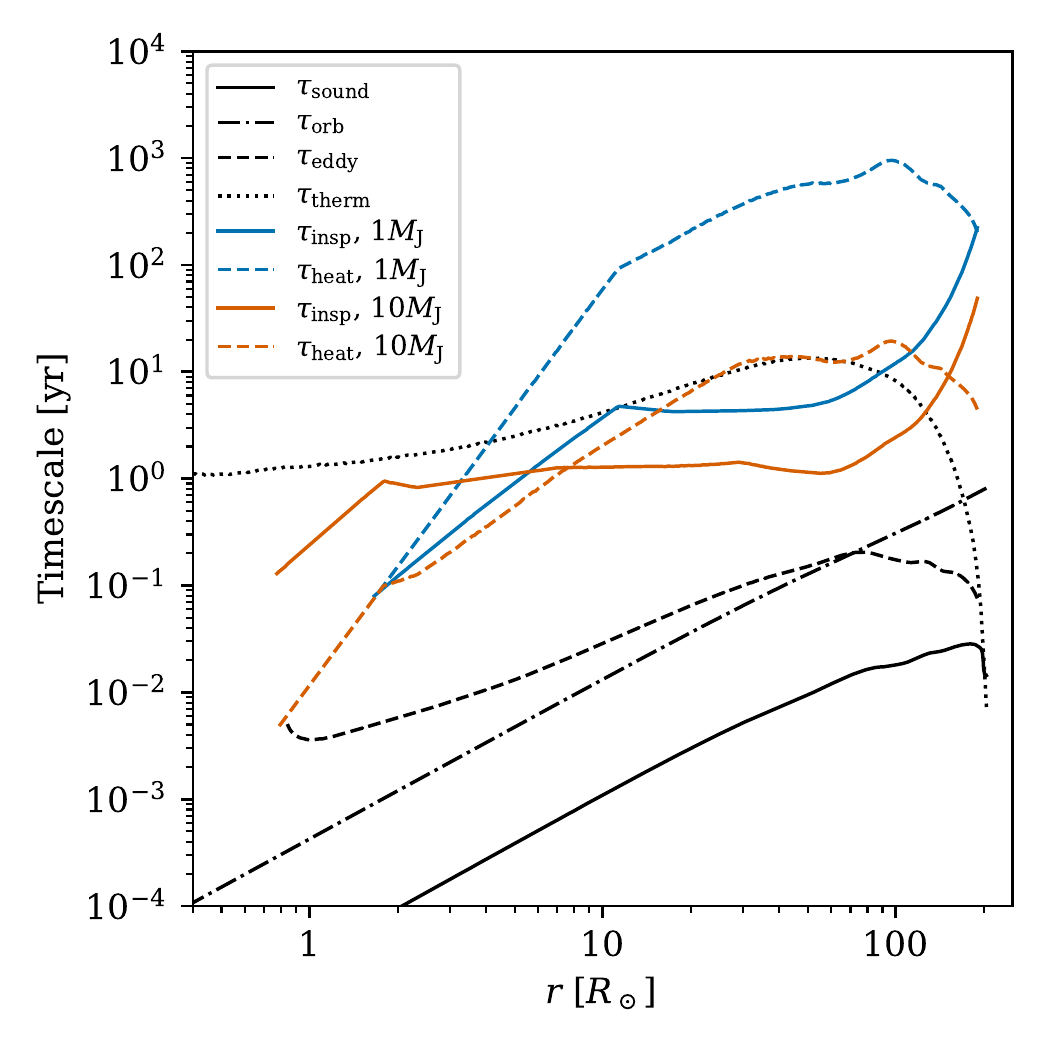}
    \caption{Characteristic timescales for $1 \MJ$ (blue) and $10 \MJ$ (vermilion) 
    planets within stellar model {\tt AGB200}. 
    Timescales that depend only on local quantities in the stellar envelope are shown in black. 
    Those that the depend on the planet's properties are in color. 
    Note the locations where the curves of $\tau_{\rm heat}$ and $\tau_{\rm therm}$ cross. 
    The bottom of the stellar convection zone is at $r \approx 0.8 \RSol$.}
    \label{fig:timescales_agb200_1mj10mj}
\end{figure}

We expect the shellular approximation to hold provided that $\tau_{\rm insp}$ is long compared to $\tau_{\rm eddy}$ and 
the size of the shock-heated region is smaller than $H_{P}$. 
If the latter is not satisfied, then a reasonable correction 
is to use $R_{\rm sh}$ as the shell thickness in lieu of $H_{P}$.

The validity of the shellular approximation is fundamental 
to our ability to study the effects of an engulfed planet using spherical 1D hydrodynamics. 
Hence, we examine the run of these relevant local timescales 
during inspiral under the passive-envelope approximation. 
Figure \ref{fig:timescales_agb200_1mj10mj} for model 
{\tt AGB200} with planets of $M_{\rm p} = 1 \MJ$ and $10 \MJ$. 
In both cases, the hierarchy of timescales necessary for the shellular approximation, 
$\tau_{\rm orb} \lesssim \tau_{\rm eddy} \lesssim \tau_{\rm insp}$, 
is satisfied for $r \lesssim 100 \RSol$. 
At greater radii, the approximation is not strictly valid 
because the convective turnover time is shorter than the orbital period. 
However, negligible heat deposition occurs in this region. 

The main differences between the $1 \MJ$ and $10 \MJ$ cases are the locations 
where the planet's gravitational focusing effect becomes unimportant 
(and thus the drag force regime changes) 
and where heating causes thermal expansion of the surroundings.
The former can be seen as the abrupt change in the slope of $\tau_{\rm insp}$ 
at $r \approx 10 \RSol$ and $r \approx 2 \RSol$, respectively. 
The latter occurs roughly where the curves of $\tau_{\rm heat}$ and $\tau_{\rm therm}$ cross, 
at $r \approx 4 \RSol$ and $r \approx 70 \RSol$, respectively. 
In the late stages of inspiral, thermal expansion occurs rapidly 
compared to the inspiral time in a passive envelope ($\tau_{\rm heat} \ll \tau_{\rm insp}$). 
This signals the breakdown of the passive-envelope approximation for massive bodies.

If $\tau_{\rm heat}$ is ever shorter than $\tau_{\rm eddy}$, 
a time-dependent treatment of convection is warranted. 
We have therefore included time-dependent convection (TDC) 
in our {\tt MESA} experiments (see \citealt{Jermyn+2022} and references therein). 
Intermittent local disruption of convection can occur for planets with $M_{\rm p} \gtrsim 5 \MJ$, 
but the global stellar response does not change significantly as a result.

Another minor complication is that evolved giant stars possess 
an inner radiative zone between the nuclear-burning shell 
and the base of the convective envelope. 
This zone has negligible mass compared to the core, 
but its radius ($r \approx 0.8 \RSol$) is such that a massive planet ($\approx 10 \MJ$) 
will interact with it before filling its Roche lobe. 
We continue to use the shellular approximation in such cases for simplicity. 

\section{{\tt MESA} simulations} \label{s:MESA}

We now describe our implementation of the effects of an engulfed planet in {\tt MESA} 
and the main results of our numerical experiments. 
The star's response to engulfment has two main regimes: 
a ``quasistatic'' regime (Section \ref{s:MESA:Quasistatic}), 
where the star remains nearly in hydrostatic equilibrium with subsonic flows; 
and a ``disruptive'' regime (Section \ref{s:MESA:Disruptive}), where it is far from equilibrium 
with supersonic flows. 

\subsection{Software information}

The {\tt MESA} equation of state (EOS) is a blend 
of the OPAL \citep{Rogers2002}, SCVH \citep{Saumon1995}, FreeEOS \citep{Irwin2004}, 
HELM \citep{Timmes2000}, PC \citep{Potekhin2010}, and Skye \citep{Jermyn2021} EOSs. 
Radiative opacities are primarily from OPAL \citep{Iglesias1993, Iglesias1996}, 
with low-temperature data from \citet{Ferguson2005}. 
Electron conduction opacities are from \citet{Cassisi2007} and \citet{Blouin2020}. 
Nuclear reaction rates are from JINA REACLIB \citep{Cyburt2010}, NACRE \citep{Angulo1999} 
and additional tabulated weak reaction rates \citep{Fuller1985, Oda1994, Langanke2000}. 
Screening is included via the prescription of \citet{Chugunov2007}. 
Thermal neutrino loss rates are from \citet{Itoh1996}.
The {\tt inlist} and extension files required to reproduce our results are available online.\footnote{\doi{10.5281/zenodo.7692746}}

\begin{figure*}
    \centering
    \begin{tabular}{cc}
        \includegraphics[width=0.5\textwidth]{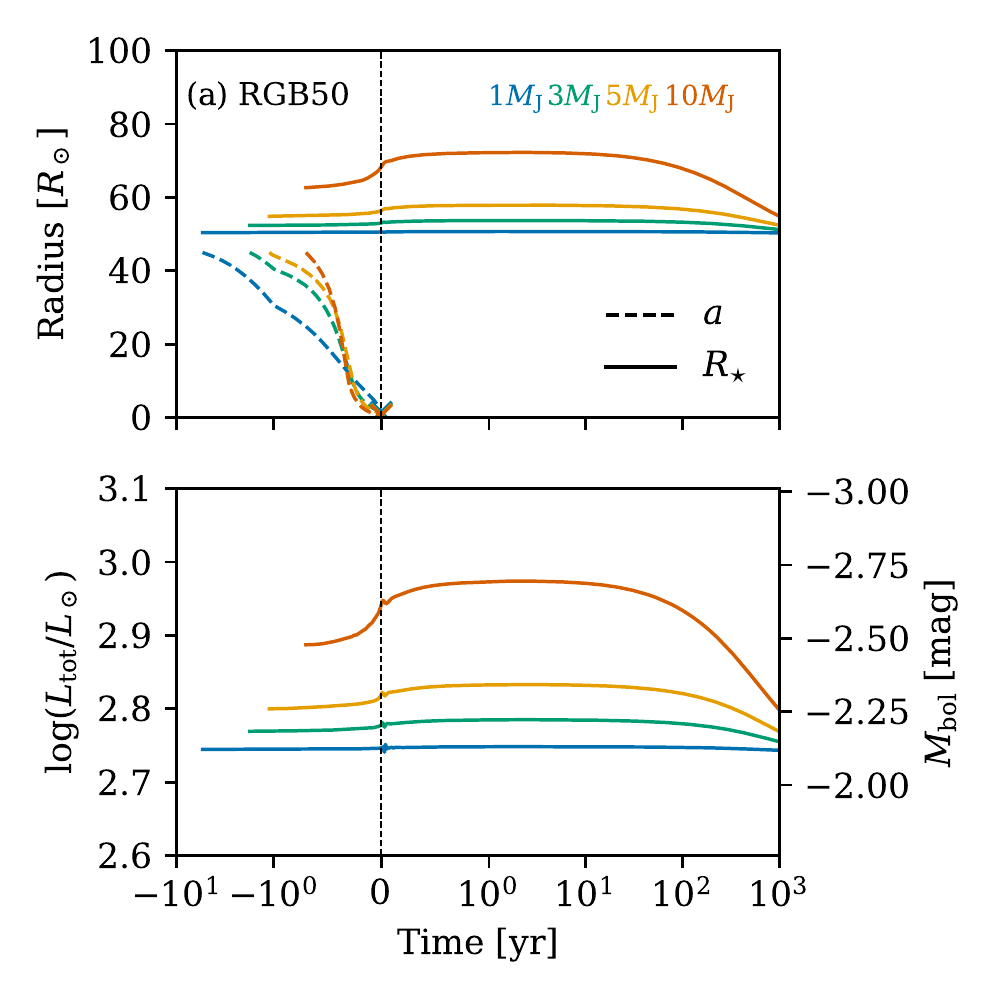} & \includegraphics[width=0.5\textwidth]{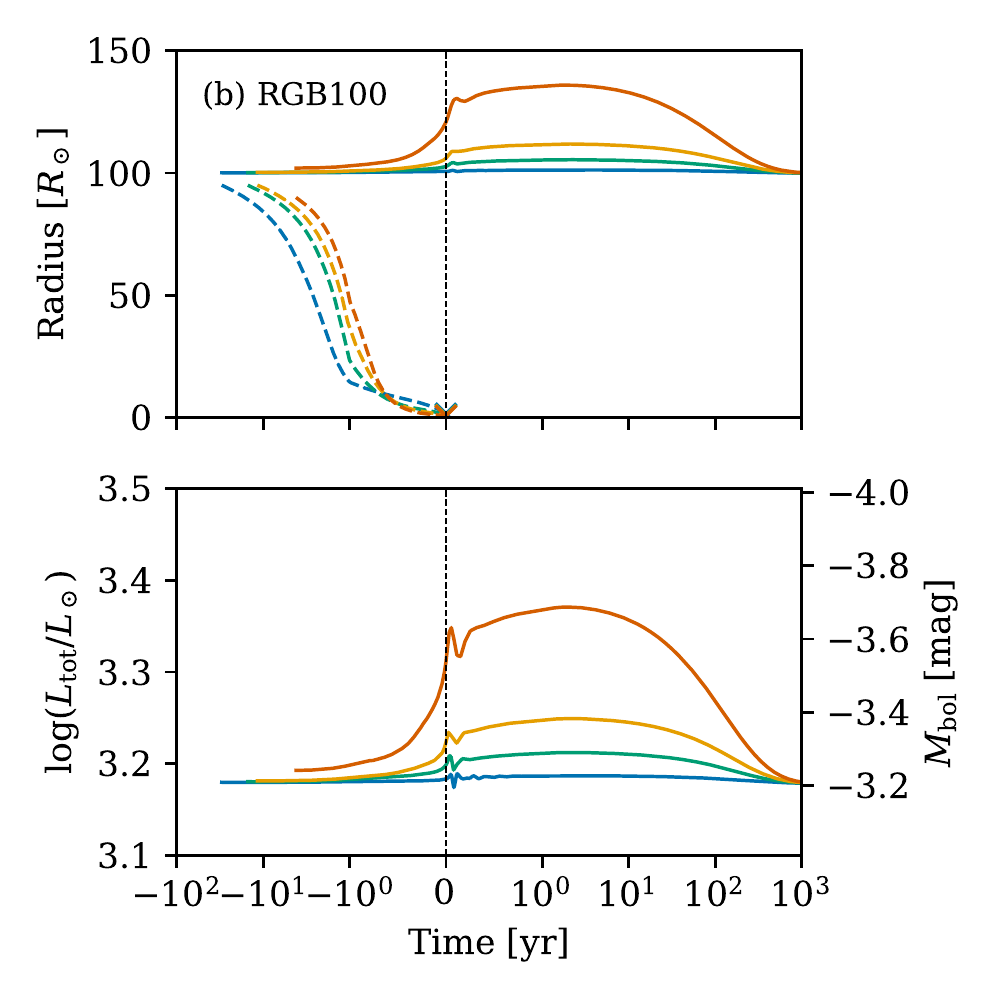} \\
        \includegraphics[width=0.5\textwidth]{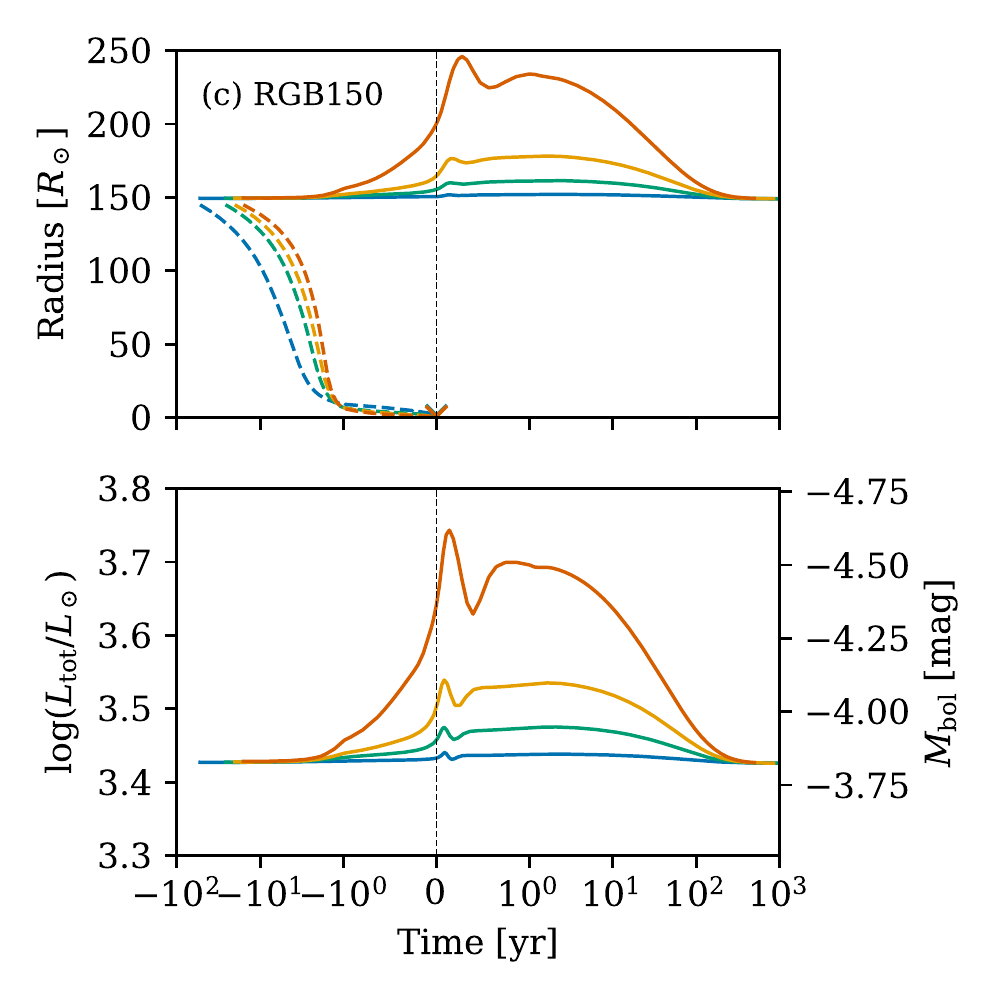} & \includegraphics[width=0.5\textwidth]{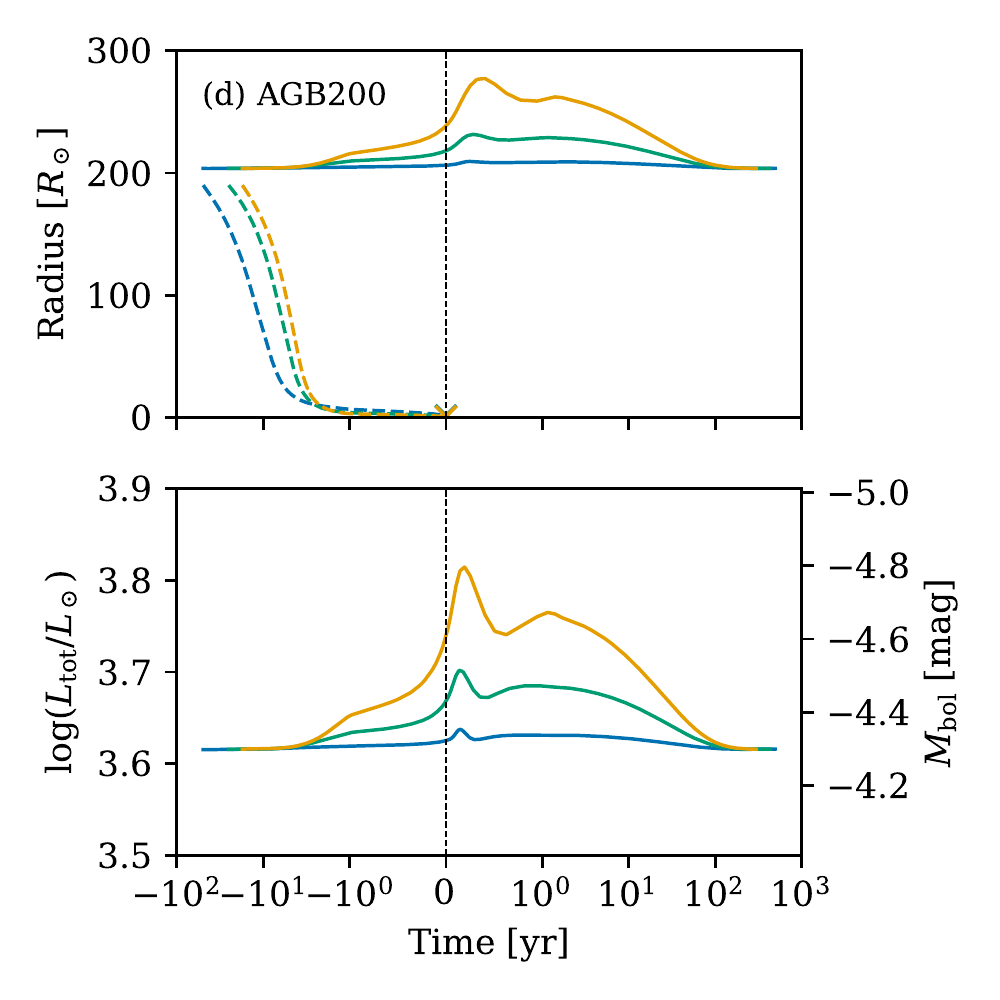} 
    \end{tabular}
    \caption{{\tt MESA} simulation results for planetary engulfment with stellar models 
    (a) {\tt RGB50}, (b) {\tt RGB100}, (c) {\tt RGB150}, (d) {\tt AGB200}. 
    Different colors correspond to different planet masses 
    [see annotations in panel (a)]. 
    Upper panels show the radii of the stellar photosphere (solid curves) and planetary orbit (dashed). 
    Lower panels show the logarithm of the star's bolometric luminosity (left abscissa) 
    and the corresponding bolometric absolute magnitude (right). 
    The ordinate shows the time before and after the planet's tidal disruption ($t=0$, vertical dotted line), 
    with a linear scale for $|t| \leq 1 \yr$ and a logarithmic scale for $|t| > 1 \yr$.}
    \label{fig:LightCurves_grid}
\end{figure*}

\begin{figure}
    \centering
    \includegraphics[width=\columnwidth]{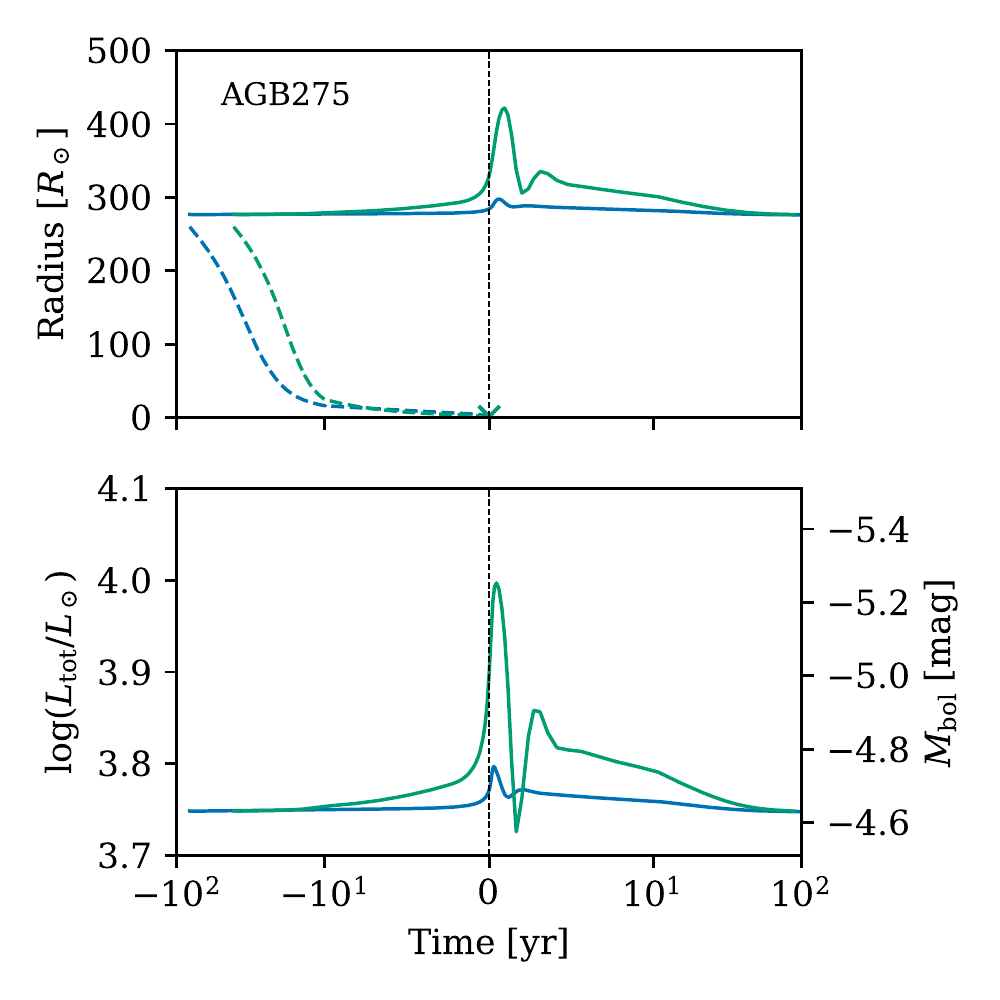}
    \caption{The same as Fig.\ \ref{fig:LightCurves_grid}, but for experiments {\tt AGB275-1MJ} and {\tt AGB275-3MJ}. 
    Here the ordinate uses a linear scale for $|t| < 10 \yr$.}
    \label{fig:LightCurves_agb275_1-3MJ}
\end{figure}

\subsection{Methods} \label{s:MESA:Methods}

In all runs, we enable {\tt MESA}'s 1D hydrodynamics module 
and treated convection using the recently added TDC option \citep{Jermyn+2022}. 
We include localized heating from an engulfed planet using the drag-force formula from Section \ref{s:InspPhysics} 
and the shellular approximation described in Section \ref{s:StellarResponse}. 
We evolve the the planetary orbit simultaneously with the stellar model 
using Eq.\ (\ref{eq:adot}) with the initial $a = 0.95 R_{\star}$. 
We use Eq.\ (\ref{eq:def_adis}) to determine when the planet is tidally disrupted 
and switch off the extra heating immediately thereafter. 
In order to focus on the hydrodynamical effects of inspiral heating, 
we do not allow for wind-driven mass loss during these simulations. 

By default, we terminated each {\tt MESA} run after an elapsed time 
$\approx 10 \tau_{\rm KH}$ since the planet's disruption, 
where $\tau_{\rm KH}$ is the stellar Kelvin--Helmholtz time. 
In most cases, this is ample time for the star to revert to its original structure. 
However, a few runs ended early because convergence could not be attained 
with a reasonable simulation time-step. 
In those cases, planetary engulfment had triggered supersonic expansion of the envelope (see Section \ref{s:MESA:Disruptive}).

We refer to our {\tt MESA} experiments using an abbreviated notation: 
for example, {\tt RGB150-3MJ} refers to the simulation 
with stellar model {\tt RGB150} (see Table \ref{tab:1.5Msun_models}) and a $3 \MJ$ planet.

\subsection{Results: Quasistatic envelopes} \label{s:MESA:Quasistatic}

Figures \ref{fig:LightCurves_grid}abc
show the simultaneous evolution of the planet and stellar structure with stellar models 
{\tt RGB50}, {\tt RGB100}, and {\tt RGB150} for $M_{\rm p} = 1$--$10 \MJ$. 
Fig.\ \ref{fig:LightCurves_grid}d shows {\tt AGB200-1MJ}, {\tt 3MJ}, and {\tt 5MJ}; 
and Fig.\ \ref{fig:LightCurves_agb275_1-3MJ} shows {\tt AGB275-1MJ} and {\tt 3MJ}. 
Each shows the evolution of the planetary orbital radius, stellar radius, and stellar bolometric luminosity and absolute magnitude. 
To facilitate comparison between simulations, 
we define the time at which the planet is tidally disrupted to be $t=0$, 
with negative and positive values of $t$ corresponding to the time before and after disruption.

The inspiral of a massive planet triggers expansion and brightening of the envelope, 
followed by protracted Kelvin--Helmholtz contraction and gradual dimming. 
The more massive the planet, the larger the disturbance it creates. 
For $M_{\rm p} = 3$--$10 \MJ$, the star's initial brightening 
is plausibly detectable with a ground-based optical or near-infrared telescope, 
with typical photometric deviations of $> 0.1$ mag over $\approx 1 \yr$. 
The effective temperature $T_{\rm eff}$ also decreases by $\sim 10$--$500 \, {\rm K}$ as the star expands, 
so a mild amount of intrinsic reddening is expected. 
We emphasize that the brightening is powered by the planet's descent in the deep interior, 
not by a `grazing' interaction at the stellar surface. 
A star may not brighten noticeably for decades to centuries after engulfment begins.

The expected shape of the light curve in this phase depends mostly 
on the host star's evolutionary stage: 
compact, early-stage red giants such as {\tt RGB50} and {\tt RGB100} 
tend to brighten in a monotonic fashion before leveling off. 
The puffy, more evolved {\tt RGB150}, {\tt AGB200}, and {\tt AGB275} 
display two distinct peaks in their light curves separated by $\approx 1 \yr$. 
Smaller secondary peaks can also be seen for {\tt RGB50} and {\tt RGB100}. 
We discuss the physical origin of the double-peaked light curve below.

The results for {\tt AGB275} are notable because the star's convective envelope
has $E_{\rm tot} > 0$, perhaps making it more susceptible to ejection \citep{PZ1968}.
Although planets of $1 \MJ$ and $3 \MJ$ cause relatively large disturbances, 
they do not unbind the envelope, likely because of efficient radiative cooling 
in the expanding outer layers (see below). 

\subsection{Results: Disrupted envelopes}
\label{s:MESA:Disruptive}

We now examine the subset of our {\tt MESA} experiments 
in which planetary engulfment disrupts the stellar envelope,  
namely {\tt AGB200-10MJ}, {\tt AGB275-5MJ}, and {\tt AGB275-10MJ}.
These represent a transitional regime between the quasistatic and subsonic responses described above
and the dynamical ejection of matter during a true common-envelope event.

\begin{figure}
    \centering
    \includegraphics[width=\columnwidth]{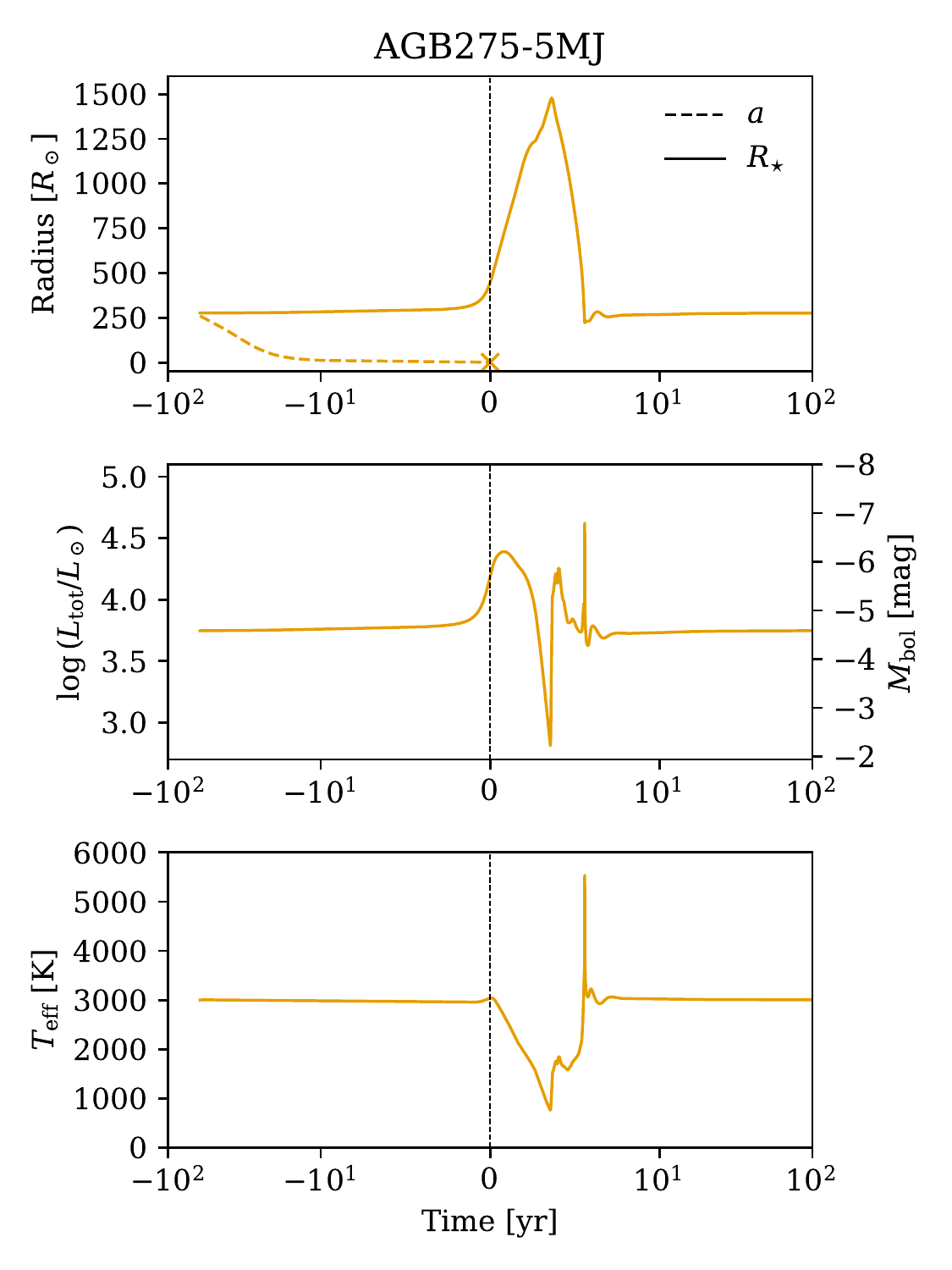}
    \caption{Top and middle panels: Similar to Fig.\ \ref{fig:LightCurves_agb275_1-3MJ} for experiment {\tt AGB275-5MJ}. The ordinate has a linear scale for $|t| < 10 \yr$. Lower panel: Evolution of the star's effective temperature.}
    \label{fig:LightCurves_agb275_5mj_new}
\end{figure}

Figure \ref{fig:LightCurves_agb275_5mj_new} shows the results of experiment {\tt AGB275-5MJ}, including the evolution of $T_{\rm eff}$. 
The evolution of the stellar radius resembles a more extreme version of experiment {\tt AGB275-3MJ} 
in that the star undergoes large-scale expansion and contraction, reaching a maximum radius of $1500 \RSol$ ($\approx 7 \AU$). 
However, this representation obscures the rich hydrodynamics of interior mass shells (see below). 
The light curve and $T_{\rm eff}$ evolution reflect this complexity more effectively. 
The star initially brightens by $\approx 1.5$ mag over $\approx 2 \yr$ before fading by $\approx 4$ mag on a similar timescale. 
At the same time, it reddens from $T_{\rm eff} \approx 3000 \,{\rm K}$ to just under $1000 \, {\rm K}$. 
When the expansion reverses, the star abruptly returns almost to the same peak luminosity 
and $T_{\rm eff} \approx 2000 \, {\rm K}$. 
Unlike in the previous experiments, the star's outer layers are essentially in free-fall at this stage. 
The luminosity at $t \gtrsim 4 \yr$ is generated by an accretion shock 
at the interface between the expanding/collapsing `ejecta' and the quasistatic interior, 
rather than by residual heat from the planetary inspiral. 

The final `spike' in the light curve, accompanied by an increase in $T_{\rm eff}$ to $\approx 5500 \, {\rm K}$, 
coincides with the re-accretion of the photosphere 
and is analogous to a shock breakout in a stellar explosion. 
Its duration is $d / v_{\rm ff}$, where $v_{\rm ff}$ is the free-fall velocity at the shock
and $d$ is the thickness of an outer shell of optical depth $\bar{\tau}_{r} = c/v_{\rm ff} \sim 10^{4}$.

\begin{figure}
    \centering
    \includegraphics[width=\columnwidth]{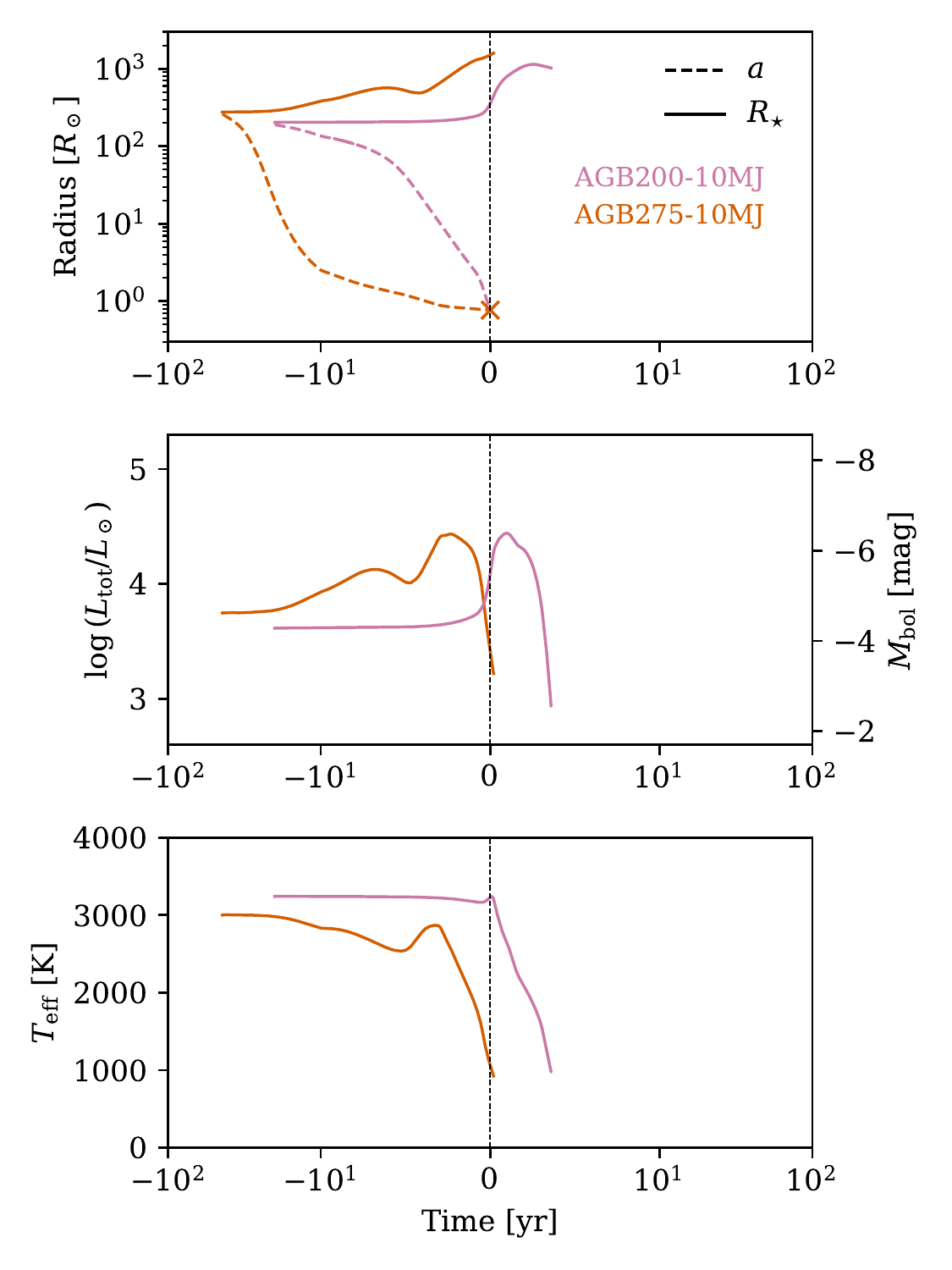}
    \caption{The same as Fig.\ \ref{fig:LightCurves_agb275_5mj_new} for experiments {\tt AGB200-10MJ} (lilac) and {\tt AGB275-10MJ} (vermilion). 
    Note that the top panel has a logarithmic vertical axis 
    to emphasize the planets' orbital evolution at small radii.}
    \label{fig:history_agb200-275_10mj}
\end{figure}

Despite the drastic expansion of the envelope, no material is ejected from the star above escape velocity. 
We note that {\tt MESA}'s equation of state takes the ionization energy of hydrogen and helium into account. 
This extra energy reservoir is evidently insufficient to eject the envelope in our scenario. 
However, we have not accounted for the formation of dust grains as the envelope expands and cools. 
Dust-driven winds may play a crucial role in the late stages of a common-envelope event \citep{GlanzPerets2018}.
If the grain opacity is large enough during the envelope's initial expansion, 
it may reduce radiative losses and allow the outer layers to reach escape velocity. 
The evolution predicted by {\tt MESA} at late times ($t \gtrsim 5 \yr$) therefore should be regarded with caution.

\markchange{Other effects not included in our simulations, 
such as stellar rotation and magnetic activity, 
would likely influence the evolution of a disrupted, dusty stellar envelope. 
If the spinning-up of the envelope by the planet generates a magnetic dynamo, 
then magnetic `cool spots' on the stellar surface may form dust grains more readily; 
this would likely change the geometry and mass-loss rate of a dust-driven outflow 
\citep[e.g.][]{Soker1998, LS2002, NordhausBlackman2006, Rapoport+2021}.}

Figure \ref{fig:history_agb200-275_10mj} shows {\tt AGB200-10MJ} and {\tt AGB275-10MJ}. 
In these cases, the {\tt MESA} runs were terminated early 
because the required time-step for convergence became prohibitively short. 
Consequently, we cannot characterize the ejecta dynamics or light curve at late times in these cases.
However, each model was evolved through the planet's tidal disruption with satisfactory accuracy. 
Broadly speaking, the star's initial expansion phase resembles that of {\tt AGB275-5MJ} in both cases. 

\subsubsection{Hydrodynamics of the interior}

\begin{figure*}
    \centering
    \includegraphics[width=\textwidth]{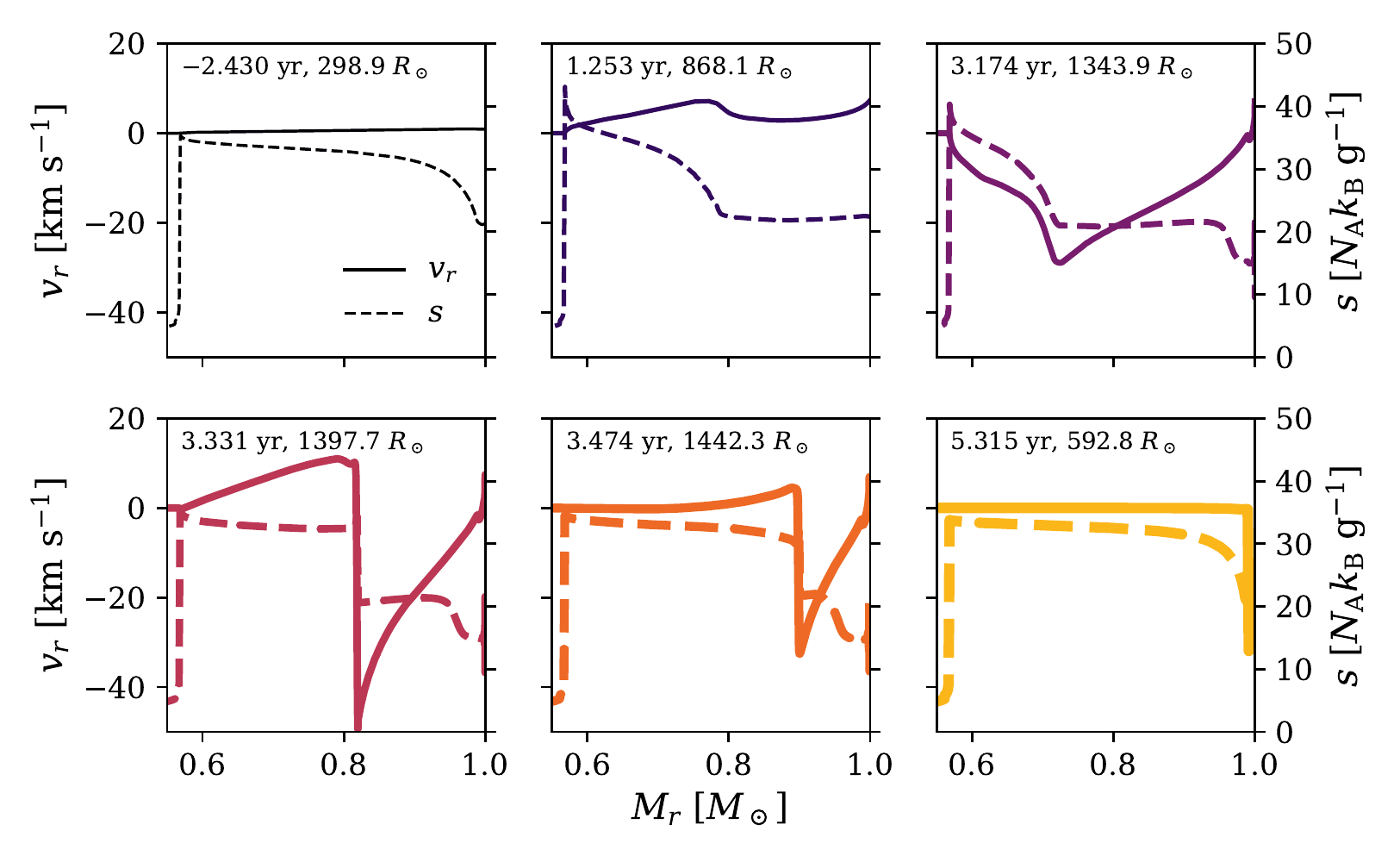}
    \caption{Snapshots of the fluid radial velocity (solid curves, left vertical axis) 
    and specific entropy (dashed, right vertical axis) 
    versus the enclosed mass coordinate $M_{r}$ for {\tt AGB275-5MJ}. 
    Each panel is annotated with the corresponding time and stellar radius. 
    The curve color and thickness in each panel matches the corresponding curves in Fig.\ \ref{fig:trad_mu_snapshots_agb275-5MJ}.}
    \label{fig:vr_s_snapshots_agb275-5MJ}
\end{figure*}

Figs.\ \ref{fig:LightCurves_agb275_5mj_new} and \ref{fig:history_agb200-275_10mj} show 
that the stellar photosphere eventually stops expanding and falls back. 
In Figure \ref{fig:vr_s_snapshots_agb275-5MJ}, we show selected snapshots 
of the radial velocity and entropy profiles of the stellar interior from {\tt AGB275-5MJ}.  
The envelope does not expand and contract monotonically throughout the evolution. 
Instead, a given mass shell undergoes multiple phases of expansion and contraction in general, 
and different mass shells can expand and contract simultaneously. 
The maximum expansion velocity is $\approx 40 \%$ of the escape speed.

\begin{figure}
    \centering
    \includegraphics[width=\columnwidth]{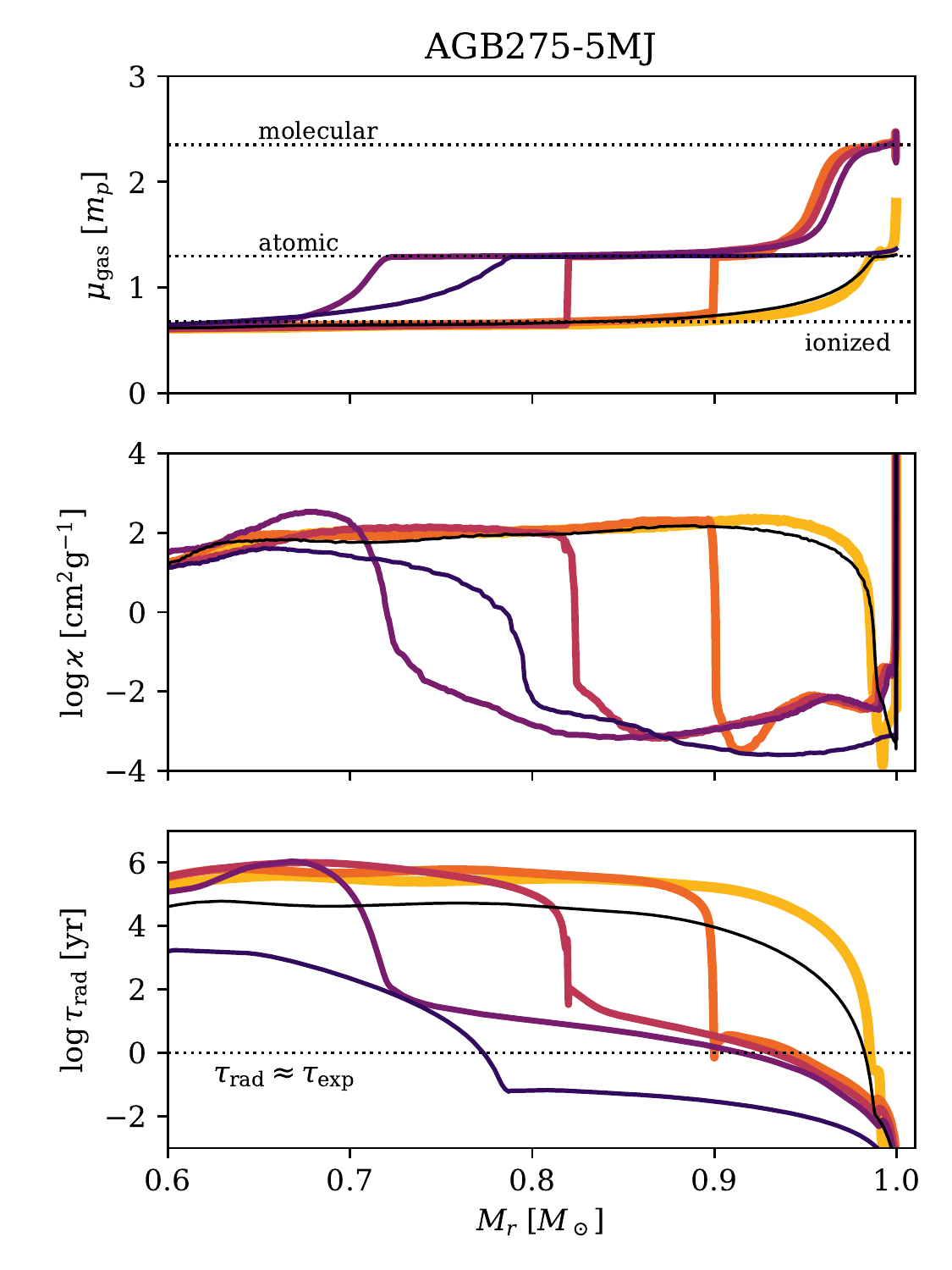}
    \caption{Snapshots of the average mass per particle (top panel), 
    radiative opacity (middle),
    and radiative cooling time (bottom) from experiment {\tt AGB275-5MJ}. 
    The color of each curve matches that in the corresponding panel of Fig.\ \ref{fig:vr_s_snapshots_agb275-5MJ}, 
    with darker/lighter colors representing earlier/later times. 
    Curves for later times are also progressively thicker.
    In the lower panel, the dotted line indicates where the $\tau_{\rm rad}$ is comparable to $\tau_{\rm exp}$, 
    i.e.\ the boundary between efficient and inefficient radiative cooling at a given time.}
    \label{fig:trad_mu_snapshots_agb275-5MJ}
\end{figure}

Initially, the star expands and contracts uniformly in the sense that the sign of the radial velocity is the same everywhere. 
The expansion is initially adiabatic, but subsequently the entropy of the outer layers 
drops abruptly as a result of hydrogen recombination (see Section \ref{s:DisruptEnv:RadiativeLosses}). 
\markchange{Radiative energy transport temporarily dominates in a large fraction of the envelope, 
due to the presence of an inverted entropy gradient (Fig.\ \ref{fig:vr_s_snapshots_agb275-5MJ}).
Radiative cooling in these regions} further deprives the expanded envelope of pressure support, leading to collapse. 
This accounts for the precipitous decline in stellar luminosity after the initial broad peak. 
However, the stellar interior eventually becomes over-pressurized from the collapse and rebounds. 
Thus, for a time, the star contains two dynamically disconnected regions: 
a subsonic expanding interior and an envelope of ejecta in supersonic free fall. 
A strong shock is present at the contact layer, 
visible as a discontinuity in the velocity and entropy profiles. 
This shock converts the kinetic energy of collapse into thermal and ionization energy, 
allowing the star to return to equilibrium.

Although the planet fails to eject the stellar envelope in our {\tt MESA} models, 
we have nonetheless identified a significant transition in their qualitative behavior: 
For very advanced stages of stellar evolution, there is a critical planetary mass above which 
the outer layers of the envelope expand supersonically and develop shocks (even if they do not become unbound). 
For {\tt AGB200} ({\tt AGB275}), the critical mass is between $5$ and $10 \MJ$ ($3$ and $5 \MJ$). 
This agrees roughly with the quantity $M_{\rm p,crit}$ 
estimated in Section \ref{s:InspPhysics:Energetics} 
via energy-budget considerations (equation \ref{eq:def_Mpcrit}). 

However, the inspiral energy budget alone cannot determine 
whether the \markchange{stellar model} undergoes a nonlinear hydrodynamical response. 
This depends rather on the rate of heat deposition during the planet's late inspiral ($L_{\rm late}$, see Section \ref{s:InspPhysics:LateInsp}), 
which is proportional to $M_{\rm p}$ for a given star. 
Assuming that the star initially undergoes quasistatic evolution, 
with $L_{\rm late}$ giving the work per unit time against self-gravity, 
the expansion rate at the surface, $v_{\rm exp}$, can be estimated as follows:
\begin{align}
    L_{\rm late} &\sim \frac{G M_{\star} M_{\rm env}}{R_{\star}^{2}} v_{\rm exp}, \nonumber \\
    \implies v_{\rm exp} &\sim \frac{L_{\rm late} R_{\star}^{2}}{G M_{\star} M_{\rm env}} \nonumber \\
    &\approx 6 \, {\rm km \, s^{-1}} \left( \frac{L_{\rm late}}{10^{5} \LSol} \right) \left( \frac{R_{\star}}{200 \RSol} \right)^{2} \nonumber \\
    & \hspace{1.5cm} \times \left( \frac{M_{\star}}{\MSol} \frac{M_{\rm env}}{0.5 \MSol} \right)^{-1}. \label{eq:v_exp_quasistatic}
\end{align}
This can be compared with the sound speed near the surface:
\begin{equation}
    c_{s} = \left( \frac{\gamma \kB T}{\mu_{\rm gas}} \right)^{1/2} \approx 12 \, {\rm km \, s^{-1}} \left( \frac{3 \gamma}{5} \frac{T}{10^{4} \, {\rm K}} \frac{m_{p}}{\mu_{\rm gas}} \right)^{1/2},
\end{equation}
where $\gamma$ and $\mu_{\rm gas}$ are the gas' adiabatic index and average mass per particle. 
In AGB stars, $v_{\rm exp}$ can be comparable to $c_{s}$ because $R_{\star}$ is large. 
If the expansion continues with $L_{\rm late}$ held constant, 
$v_{\rm exp}$ increases in proportion to $R_{\star}^{2}$ 
and $c_{s}$ drops due to adiabatic cooling. 
Consequently, the expansion becomes supersonic for large enough $L_{\rm late}$ ($\propto M_{\rm p}$), 
as we have seen. 

We have verified this picture by conducting additional {\tt AGB200-10MJ} runs 
in which the drag force on the planet is artificially reduced by a 
constant factor. 
Although the same amount of heat is deposited in the star in each case, 
the stellar response becomes quasistatic when the drag force is reduced by roughly an order of magnitude. 
This is because (a) the envelope expanded more slowly and 
(b) convection carries a greater fraction of $L_{\rm late}$.

\subsubsection{Recombination and radiative losses (without dust)} \label{s:DisruptEnv:RadiativeLosses}

Interior structure profiles reveal that a large portion of the envelope 
undergoes recombination as a result of adiabatic cooling during the star's initial expansion. 
The top panel of Figure \ref{fig:trad_mu_snapshots_agb275-5MJ} shows the evolution of $\mu_{\rm gas}$ 
throughout the envelope for experiment {\tt AGB275-5MJ} 
as a proxy for the predominant phase of hydrogen. 
As the envelope expands following the planet's inspiral, more than half of the envelope's mass undergoes recombination. 
(The temperature near the surface can be low enough to form molecules and dust grains, 
but this is a secondary effect in terms of energetics.)

In the study of CE evolution, the role of ionization energy in ejecting the envelope is uncertain
\markchange{(see discussions in the reviews of \citealt{Ivanova+2013} and \citealt{RoepkeDeMarco2022}, 
as well as e.g.\ \citealt{Sabach+2017}, \citealt{Grichener+2018}, and \citealt{Ivanova2018}). 
The crux of the issue is whether a sufficient portion of the ionization energy 
can be converted to mechanical work to accelerate the envelope to escape velocity. 
Several previous studies, including \citet{Sabach+2017}, \citet{Grichener+2018}, and \citet{WilsonNordhaus2019}, 
have argued that convection can efficiently transport heat 
derived from orbital and ionization energy from the deep interior 
to radiative zones near the surface, 
reducing the amount of energy available to do work 
(but see \citealt{Ivanova2018}).
We find that much of the envelope's initial ionization energy is lost as radiation 
during the initial large-scale expansion of the envelope (see Section \ref{s:DisruptEnv:RadiativeLosses}). 
Subsequently, the remaining heat deposited by the planet flows outward,
re-ionizining the envelope in the process. 
This implies that the fluid absorbs a substantial amount of thermal energy 
that could otherwise have been used as work to eject the envelope. 
The ionization energy therefore acts as a `buffer' for the planet's orbital energy, 
arguably preventing envelope ejection rather than aiding it. 
These findings demonstrate the nontrivial effects of ionization energy transport in CE evolution: 
even if it does not contribute to ejecting the envelope, 
it may be necessary to include ionization effects in numerical calculations 
in order to accurately predict a system's evolution and observable characteristics.
However, we reiterate that our conclusions are provisional, due to the lack of dust effects in our simulations.}

Recombination has two main effects on stellar matter in our {\tt MESA} models: 
The ionization energy converts to radiation, and the opacity plummets. 
Even though the envelope remains optically thick, 
the reduced opacity enables efficient radiative cooling. 
To demonstrate this, the middle and bottom panels of Fig.\ \ref{fig:trad_mu_snapshots_agb275-5MJ} 
show the evolution of the radiative opacity $\kappa$ reported by {\tt MESA} 
and a local radiative cooling timescale $\tau_{\rm rad}$. 
The cooling time is defined as
\begin{equation}
    \tau_{\rm rad} = \bar{\tau}_{r} \frac{R_{\star} - r}{c} \frac{P_{\rm gas}}{P_{\rm rad}},
\end{equation}
where $\bar{\tau}_{r} = \int_{r}^{R_{\star}} \kappa(r') \rho(r') \, \dif r'$ 
is the optical depth at radius $r$, $c$ is the speed of light, 
and $P_{\rm gas}$ and $P_{\rm rad}$ are the local gas pressure and radiation pressure. 
In the layers that undergo recombination, $\tau_{\rm rad}$ 
becomes comparable to, or shorter than, the envelope expansion time 
$\tau_{\rm exp} \approx 1 \yr$. 
Roughly speaking, layers with $\tau_{\rm rad} < \tau_{\rm exp}$ can cool efficiently. 

This line of reasoning also reproduces the maximum luminosity 
during the initial expansion of the envelope. 
Assuming that a shell of mass $M_{\rm sh}$ loses 
all of its hydrogen ionization energy as radiation over a time $\tau_{\rm exp}$, 
we find that the star's maximum luminosity is given by:
\begin{align}
    L_{\rm max} &= \frac{X M_{\rm sh} q_{\rm H}}{\tau_{\rm exp}} \nonumber \\
    &\approx 3.2 \times 10^{4} \LSol \left( \frac{X}{0.75} \frac{M_{\rm sh}}{0.2 \MSol} \right) \left( \frac{\tau_{\rm exp}}{1 \yr} \right)^{-1}, \label{eq:Lmax_LRN}
\end{align}
where $X$ is the hydrogen mass fraction and $q_{\rm H} = 13.6 \, {\rm eV} / m_{p}$. 
This agrees with the {\tt MESA} results for {\tt AGB200-10MJ}, {\tt AGB275-5MJ}, and {\tt AGB275-10MJ}. 
(All else being equal, accounting for the ionization energy of helium 
increases $L_{\rm max}$ by $10\%$.)
Recombination is also responsible for the first peak in the light curves 
of experiments {\tt RGB150-10MJ}, {\tt AGB200-5MJ}, and {\tt AGB275-3MJ} 
(Figs.\ \ref{fig:LightCurves_grid}cd, \ref{fig:LightCurves_agb275_1-3MJ}). 
In those cases, the envelope expands by a smaller factor and remains subsonic, 
so $M_{\rm sh}$ and $L_{\rm max}$ are reduced relative to the reference values above. 

As noted above, dust grains are expected to form as the stellar envelope expands and cools. 
Their additional opacity may reduce radiative energy losses, 
allowing more efficient ejection of the envelope. 
In order for this to occur, the grain opacity must be large enough  
that $\tau_{\rm rad}$ remains longer than $\tau_{\rm exp}$. 
Referring to the bottom panel of Fig.\ \ref{fig:trad_mu_snapshots_agb275-5MJ}, 
we see that the grain opacity must be 2--3 orders of magnitude greater than the low-temperature {\tt MESA} opacity 
to eliminate radiative losses completely in {\tt AGB275-5MJ}. 

\section{Observational implications} \label{s:Observations}

\subsection{Optical and infrared transients}

Previous studies have suggested that planet engulfment \markchange{(or tidal disruption) 
may lead to an observable optical/infrared transient, 
including \citet{RM2003}, \citet{Retter+2006, Retter+2007}, \citet{Bear+2011}, 
\citet{Metzger+2012, Metzger+2017}, \citet{KashiSoker2017}, \citet{Soker2018}, 
\citet{MacLeod2018}, \citet{Kashi+2019}, \citet{Stephan+2020}, and \citet{Gurevich+2022}.
}
Our results confirm this prediction and characterize the relationship 
between transient properties and underlying planetary and stellar parameters. 

\begin{figure}
    \centering
    \includegraphics[width=\columnwidth]{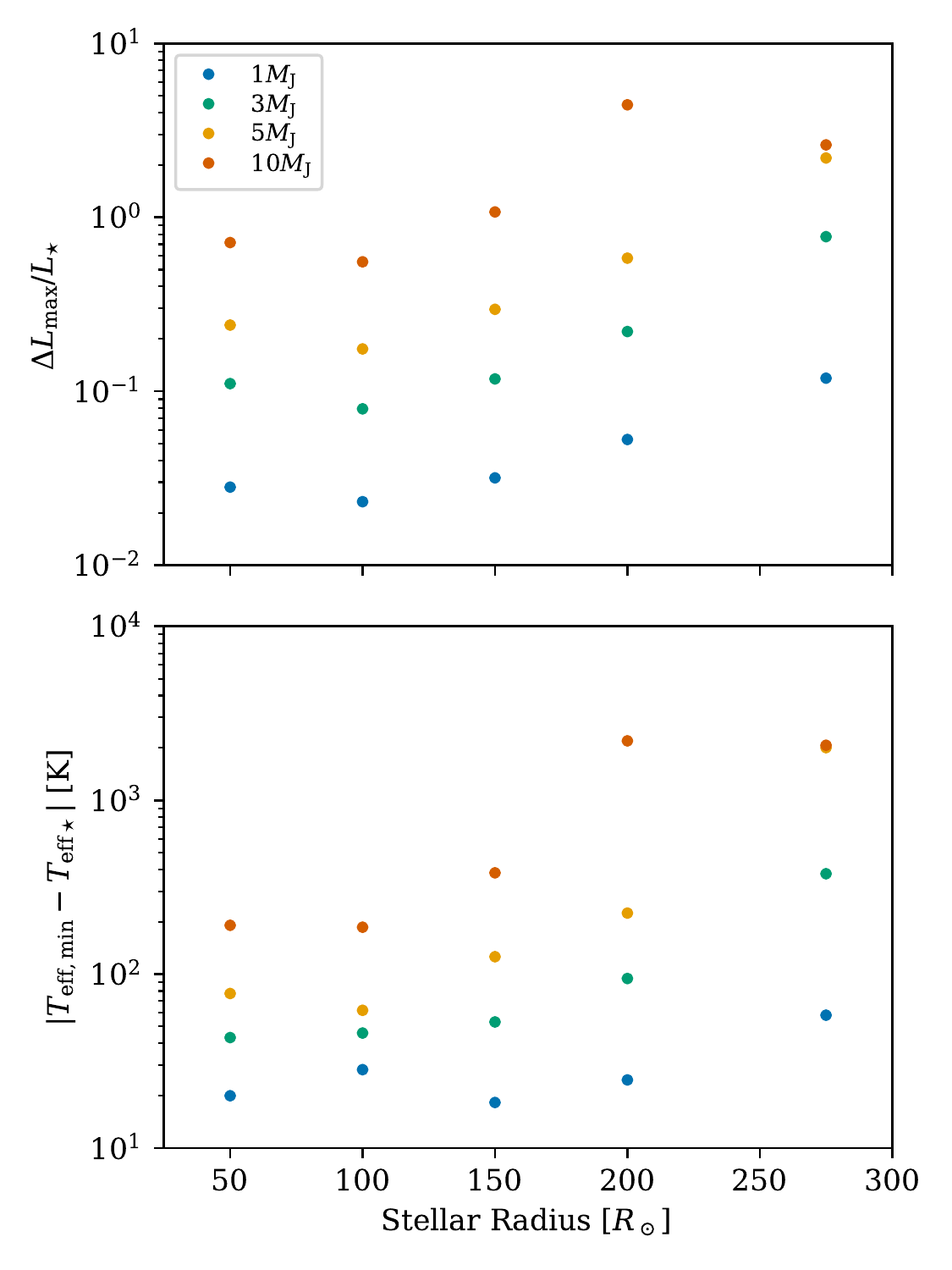}
    \caption{Maximum fractional change of the stellar bolometric luminosity (upper panel) 
    and maximum change of the effective temperature (lower) 
    for various combinations of stellar models and planetary masses.}
    \label{fig:MESA_DLmaxRstar}
\end{figure}

The most informative quantity to constrain the engulfed planet's mass is the excess stellar luminosity. 
Let $\Delta L(t)$ be the difference between the star's instantaneous bolometric luminosity 
$L_{\rm tot}(t)$ and its intrinsic luminosity $L_{\star}$. 
Let the maximum value be $\Delta L_{\rm max}$ occurring at time $t_{\rm max}$.
The upper panel of Figure \ref{fig:MESA_DLmaxRstar} shows the quantity 
$\Delta L_{\rm max} / L_{\star}$ for our {\tt MESA} simulations. 
For stars that experience quasistatic or subsonic responses 
(which generally correspond to $\Delta L_{\rm max} / L_{\star} \lesssim 1$), 
it may be possible to estimate the planetary mass from the peak luminosity, 
provided that the host's properties are well constrained prior to the flare-up. 
However, when the planet creates a major disturbance in the star (Section \ref{s:MESA:Disruptive}), 
non-linear effects make the relation between $M_{\rm p}$ and $\Delta L_{\rm max}$ more complex.

The maximum change of the host's intrinsic color 
(which we characterize via $T_{\rm eff}$) 
also correlates with planetary mass. 
The effective temperature tends to decrease somewhat during the transient, reddening the star. 
The lower panel of Fig.\ \ref{fig:MESA_DLmaxRstar} shows 
the maximum change of $T_{\rm eff}$ from its initial value. 
Cases with more massive planets and more evolved host stars exhibit larger $T_{\rm eff}$ changes. 
Again, however, non-linear effects dominate when the planet disrupts the envelope. 

\begin{figure}
    \centering
    \includegraphics[width=\columnwidth]{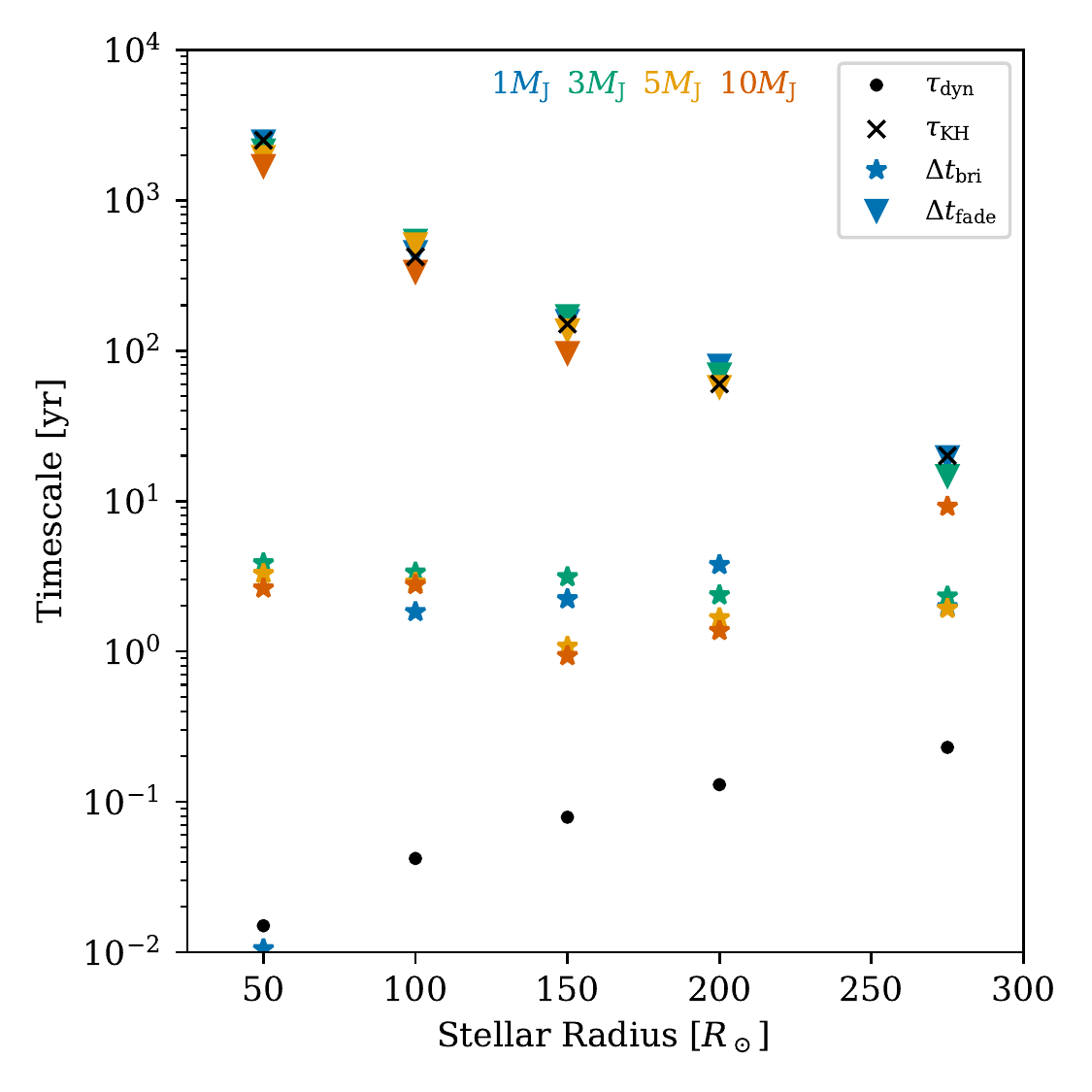}
    \caption{The brightening and fading timescales of
    the {\tt MESA} light curves are shown as star-shaped and triangular points, 
    color-coded by planetary mass. 
    The host star's global dynamical time and Kelvin--Helmholtz time 
    are also shown as black points and crosses, respectively. 
    Note $\Delta t_{\rm fade}$ is undefined for {\tt AGB200-10MJ}, {\tt AGB275-5MJ}, {\tt AGB275-10MJ}.
    }
    \label{fig:MESA_LCTimes}
\end{figure}

The brightening and fading timescales of engulfment-powered transients are also important. 
In Figure \ref{fig:MESA_LCTimes}, we show the values of the empirical quantities 
$\Delta t_{\rm bri}$ and $\Delta t_{\rm fade}$, 
defined as follows: 
$\Delta t_{\rm bri} = t_{\rm max} - t_{\rm A}$ and $\Delta t_{\rm fade} = t_{\rm B} - t_{\rm max}$, 
where $\Delta L(t_{\rm A}) = \Delta L(t_{\rm B}) = 0.1 \Delta L_{\rm max}$ and $t_{\rm A} < t_{\rm max} < t_{\rm B}$.

For systems with quasistatic or subsonic responses, the fading time matches the Kelvin--Helmholtz time,
\begin{align}
    \tau_{\rm KH} &= \frac{G M_{\star}^{2}}{R_{\star} L_{\star}} \nonumber \\
    &\approx 40 \yr \left( \frac{M_{\star}}{\MSol} \right)^{2} \left( \frac{R_{\star}}{200 \RSol} \frac{L_{\star}}{4000 \LSol} \right)^{-1}. \label{eq:def_tKH}
\end{align}
This shows that the star contracts quasistatically 
as the planet's orbital energy is lost as radiation. 
Since $\tau_{\rm KH}$ is longer than a century for stars with radii $\lesssim 100 \RSol$, 
the secular dimming of star following engulfment is measurable only for TRGB and AGB stars. 
However, it may be possible to detect the contraction of stars in the $\approx 100$--$150 \RSol$ range by indirect means (see below). 

The brightening time varies within the range 
of $\sim 1$--$5 \yr$ across almost all of our runs, 
without any apparent trend with planetary mass for a given stellar model or between stellar models. 
It is long compared to the global dynamical timescale,
\begin{align}
    \tau_{\rm dyn} &= \left( \frac{R_{\star}^{3}}{G M_{\star}} \right)^{1/2} \nonumber \\
    &\approx 0.14 \yr \left( \frac{R_{\star}}{200 \RSol} \right)^{3/2} \left( \frac{M_{\star}}{\MSol} \right)^{-1/2}, \label{eq:def_tdyn}
\end{align}
which reflects the star's expansion below escape velocity. 
If we assume that the star undergoes quasistatic expansion during the planet's late inspiral, 
we estimate an expansion timescale (see Eq.\ \ref{eq:v_exp_quasistatic})
\begin{equation} \label{eq:def_texp_static}
    \tau_{\rm exp} = \frac{G M_{\star} M_{\rm env}}{R_{\star} L_{\rm late}} \sim \tau_{\rm KH} \left( \frac{L_{\rm \star}}{L_{\rm late}} \right).
\end{equation}
Using $L_{\rm late} \approx 10^{4}$--$10^{6} \LSol$ 
for the appropriate stellar parameters (Fig.\ \ref{fig:Llate_passive}), 
Eq.\ (\ref{eq:def_texp_static}) reproduces the range of $\Delta t_{\rm bri}$ seen in our {\tt MESA} results. 
Run {\tt RGB50-1MJ} is an exception because the stellar response is so weak 
that small acoustic oscillations (with period $\approx \tau_{\rm dyn}$) 
excited during the late inspiral dominate the light curve. 

\subsubsection{Weak red transients} 

Broadly speaking, the electromagnetic signatures of planetary inspiral 
can be separated into two qualitative categories 
based on whether the stellar response was quasistatic/subsonic or supersonic. 
The main signature of a quasistatic/subsonic envelope is an abrupt ($\approx 1 \yr$) increase in luminosity with mild reddening 
(see Figs.\ \ref{fig:LightCurves_grid}, \ref{fig:LightCurves_agb275_1-3MJ} for examples). 
TRGB and AGB stars may exhibit a prominent `double peak' in bolometric luminosity due to H recombination. 
In most cases, the transient phase is followed by a long plateau ($10$--$100 \yr$) at nearly constant $L$ and $T_{\rm eff}$ 
as Kelvin--Helmholtz contraction takes over. 
The bolometric amplitudes of these transients are $\sim 0.1$--$1$ mag for planets between $3$ and $10 \MJ$. 
Despite their modest amplitudes, these ``weak red transients'' may be observable 
by ground- and space-based wide-field time-domain surveys operating at optical and near-infrared wavelengths, 
such as the Zwicky Transient Facility, the Vera C.\ Rubin Observatory, and the Nancy Grace Roman Space Telescope. 

We estimate the rate of weak red transients in the Galaxy 
based on the rate of WD formation ($\Gamma_{\rm WD} \sim 1 \yr^{-1}$) 
and the occurrence fraction of warm Jupiters around FGK dwarfs ($f_{\rm WJ} \approx 0.03$; \citealt{Cumming+2008, Mayor+2011}):
\begin{equation}
    \Gamma_{\rm WRT} = f_{\rm WJ} \Gamma_{\rm WD} \approx 0.03 \yr^{-1} \left( \frac{f_{\rm WJ}}{0.03} \frac{\Gamma_{\rm WD}}{1 \yr^{-1}} \right)
\end{equation}

\subsubsection{Red novae from engulfment on the AGB}

Despite some physical uncertainties and computational limitations, 
our {\tt MESA} experiments clearly predict that sufficiently massive planets 
can disrupt an AGB star's envelope, producing a major eruption lasting several years (Figs.\ \ref{fig:LightCurves_agb275_5mj_new}, \ref{fig:history_agb200-275_10mj}). 
These eruptions likely resemble the luminous red novae (LRNe) 
produced by coalescing binary stars \citep[e.g.][]{TylendaSoker2006}. 
However, they are much dimmer at peak brightness and evolve more slowly than typical LRNe \citep[e.g.][]{Kochanek2014, Karambelkar+2022}. 
Additionally, their progenitors are already bright, dusty, late-type sources by virtue of their evolutionary stage,  
whereas the progenitors of LRNe display a wide range of properties. 

If the fraction of WD progenitors with a cold Jupiter is $f_{\rm CJ}$, 
the rate of engulfment-powered RNe in the Galaxy is
\begin{equation}
    \Gamma_{\rm RNe} \approx 0.1 \yr^{-1} \left( \frac{f_{\rm CJ}}{0.1} \frac{\Gamma_{\rm WD}}{1 \yr^{-1}} \right),
\end{equation}
where we again estimate $f_{\rm CJ}$ based on the occurrence rate among Sun-like stars \citep{Fernandes+2019, Fulton+2021}. 
If the giant-planet occurrence rate is greater for stars somewhat more massive than the Sun 
\citep{Johnson+2010, Reffert+2015, Jones+2016, Ghezzi+2018}, 
then a somewhat larger value of $f_{\rm CJ}$ may be appropriate 
for typical single WD progenitors ($M_{\star} \approx 1.5$--$3 \MSol$ on the MS). 
The higher rate and peak luminosity of these eruptions relative to weak red transients 
makes them a more promising target population for transient surveys. 

When an AGB star is disrupted by an engulfed planet, 
it reaches a typical peak luminosity $M_{\rm bol} \approx -6$ and lasts several years. 
These properties, combined with the expected event rate $\sim 0.1 \yr^{-1}$, 
suggest that planetary engulfment events could be responsible for a significant fraction 
of the lower-luminosity red novae observed in the Galaxy \citep{Kochanek2014, Howitt+2020}. 
The transient OGLE-2002-BLG-360 \citep{Tylenda+2013}, 
which was ``less violent'' and of longer duration than a typical LRN, 
and whose progenitor was a late-type giant star, may be an example.

\subsubsection{\markchange{Observational complications}}

\markchange{Here we note some possible complications 
in associating an observed transient with the engulfment of a giant planet. 
For AGB host stars, there is potential for confusion between planet-engulfment events 
and other sources of short-term variability. 
Many AGB stars display long-period intrinsic variability, especially Mira-like pulsations. 
The typical pulsation period and photometric amplitude of a Mira variable 
are $\sim 1 \yr$ and $\gtrsim 1 \, {\rm mag}$ \citep[e.g.][]{Iwanek+2022}.
Variability at this level could mask the expansion and brightening of an AGB star 
due to engulfment of a lower-mass ($\lesssim 3 \MJ$) companion.} 

\markchange{AGB stars also experience brief, recurrent increases in their luminosity 
and mass-loss rate due to helium shell flashes (a.k.a.\ thermal pulses). 
These events have been associated with the creation of detached shells of gas and dust 
observed around some nearby AGB stars 
\citep[e.g.][]{Olofsson+1988, Maercker+2016, Kerschbaum+2017, Brunner+2019, KastnerWilson2021}. 
Detached shells have a visible lifetime of $\sim 3 \times 10^{4} \yr$ \citep{KastnerWilson2021}. 
A few stars display multiple shells \citep{Izumiura+1997, Mecina+2014}, 
each presumably resulting from a distinct mass-loss episode. 
Engulfment of a giant planet could also create a detached shell, 
for a large enough increase of the stellar luminosity. }

\markchange{The observational consequences of a thermal pulse in an AGB star 
may be broadly similar to those of planetary engulfment. 
However, in practice, we do not expect thermal pulses to contaminate searches 
for transients caused by planetary engulfment. 
This is because the duration of elevated luminosity and mass loss 
resulting from a helium shell flash ($\sim 1000 \yr$) 
is much longer than that of a red nova due to an engulfment event in our models ($\sim 1$--$10 \yr$).}

\markchange{Finally, we noted that the maximum luminosity and $T_{\rm eff}$ change
do not correlate with $M_{\rm p}$ for AGB stars engulfing $5$--$10 \MJ$ planets (Fig.\ \ref{fig:MESA_DLmaxRstar}). 
This raises the question of whether one could distinguish 
between the engulfment of a $5$--$10 \MJ$ planet by an AGB star 
and the engulfment of a brown dwarf ($10$--$80 \MJ$) 
or low-mass M dwarf ($30$--$200 \MJ$). 
We have argued that the maximum luminosity of a disrupted AGB envelope 
scales with $M_{\rm sh} / \tau_{\rm exp}$ (see Eq.\ \ref{eq:Lmax_LRN} and surrouding text). 
Observational evidence shows that a companion $\gtrsim 50 \MJ$ can survive the CE phase 
\citep[e.g.][]{Maxted+2006, vanRoestel+2021, Kruckow+2021, ZS2022}. 
The inspiral of such a companion proceeds on the global dynamical timescale 
(Eq.\ \ref{eq:tinsp_grav} for $M_{\rm p} \sim M_{\star}$) 
and results in a complete envelope ejection. 
Eq.\ (\ref{eq:Lmax_LRN}) predicts a much more luminous eruption, 
assuming $\tau_{\rm exp} \ll 1 \yr$ and $M_{\rm sh} \simeq M_{\rm env}$. 
If the companion stalls on a short-period orbit, then the residual envelope 
may experience additional eruptions on a timescale of years \citep{Clayton+2017}.
In the intermediate regime of $10$--$50 \MJ$, 
it is unclear whether $M_{\rm sh}$ or $\tau_{\rm exp}$ would change significantly. 
We have not attempted to simulate this regime in {\tt MESA} 
because the conditions for our 1D spherical approximation 
may not be met in that case. 
Further study is required to determine to what extent 
giant-planet and low-mass brown-dwarf engulfment events are distinguishable.}

\subsection{Seismology of post-engulfment giants}

In the `weak red transient' regime, the dimming timescale following planet engulfment 
is longer than the brightening timescale by a factor of $\sim 10$--$10^{3}$, depending on evolutionary stage. 
Thus, there are $\sim 10$--$10^{3}$ times more stars in the dimming phase 
than the brightening phase at any given time. 
For the most part, the dimming is too gradual 
to be measured on human timescales. 
However, it may be possible to detect the star's quasistatic contraction using asteroseismology. 
Specifically, a contracting star would display a secular increasing trend 
in the large seismic frequency spacing $\Delta \nu \propto 1/\tau_{\rm dyn} \propto R_{\star}^{-3/2}$. 
The {\it Kepler} mission produced a homogeneous set of $\Delta \nu$ measurements 
with a typical fractional uncertainty of $10^{-3}$
for several thousand RGB stars monitored continuously over $\approx 4 \yr$ \citep{Yu+2018}. 
Stars contracting at rates $|\dot{R}_{\star}| / R_{\star} \gtrsim 10^{-3} \yr^{-1}$
may have a detectable $\Delta\nu$ trend. 

Additionally, giant stars that have engulfed a giant planet rotate more rapidly 
than is typically observed or expected 
based on theoretical models of single stellar evolution
(see Section \ref{s:InspPhysics:OrbDecayHeat:SpinUp} and references therein). 
\markchange{Rapid rotation would cause rotational mode splitting in the seismic spectrum 
and would be a longer-lasting signature of planet engulfment than brightness changes.
The spin-down timescale of a red giant 
due to stellar mass loss 
is $\simeq 0.1 (M_{\star} / \dot{M}_{\rm w}) ( R_{\star} / r_{\rm w})^{2}$,
where $\dot{M}_{\rm w}$ is the rate of mass loss due to a stellar wind 
and $r_{\rm w}$ is the radius from which the wind is launched. 
For $r_{\rm w} \approx R_{\star}$, the spin-down time is $\approx 10\%$
of the star's remaining lifetime ($\sim M_{\star}/\dot{M}_{\rm w}$).}
A giant displaying both strong mode splitting 
and upward secular drift of $\Delta \nu$ 
would be a strong candidate for having recently engulfed a substellar companion. 

\markchange{
For completeness, we note that an engulfed planet or brown dwarf can excite stellar oscillations directly 
during its inspiral phase \citep[e.g.][]{Soker1992a, Soker1992b, GagnierPejcha2023}. 
This has possible ramifications for wind-driven mass loss 
and the detectability of ongoing engulfment events.
}

\subsection{Planetary survival}

As mentioned previously, this work is motivated in part by the question 
of whether an engulfed giant planet can survive on a short-period orbit around a WD 
after ejecting the stellar envelope 
\citep{Vanderburg+2020, Lagos+2021, Chamandy+2021, Merlov+2021}. 
In all of our {\tt MESA} experiments, the planet undergoes Roche-lobe overflow (RLO) before the end of the simulation. 
We have assumed that, once RLO begins, the planet is disrupted rapidly 
compared to the ongoing dynamical evolution of the envelope. 
This assumption bears further scrutiny. 

In principle, RLO of a short-period giant planet can lead to stable mass transfer, 
halting or even reversing its orbital decay \citep{Valsecchi+2014,Jackson+2016,JS2017}. 
In our case, the inspiral/mass-transfer timescale 
is much shorter than the planet's thermal timescale. 
The planet's radius therefore evolves at a fixed entropy per unit mass, 
increasing or remaining nearly constant \citep[e.g.][]{ZS1969, Paxton+2013}. 
Thus, the mass transfer is unstable and runaway disruption ensues. 

RLO can be prevented if the stellar envelope expands to such a degree 
that the planet's inspiral time becomes longer than the global dynamical time, 
`stalling' the planet at $a > a_{\rm dis}$ until the envelope has dissipated completely. 
In `successful' CE ejections, these self-regulating behaviors prevent a complete merger 
and set the post-CE binary separation 
\citep[e.g.][]{Ivanova+2013, Clayton+2017, GagnierPejcha2023}. 
The only experiment in which the inspiral became somewhat self-regulating was {\tt AGB275-10MJ} (Fig.\ \ref{fig:history_agb200-275_10mj}). 
However, this merely delayed RLO by $\approx 10 \yr$, 
much less than the expected timescale of dust-driven mass loss 
from an AGB or post-CE envelope \citep[e.g.][]{GlanzPerets2018}. 
This suggests that planetary survival is possible only for $M_{\rm p} \gtrsim 10 \MJ$
and only when engulfment occurs during the thermally pulsing AGB stage. 
This, in turn, implies that short-period giant planets can only be found orbiting WDs with carbon/oxygen cores 
(as opposed to close `WD + brown dwarf' binaries, which often contain helium-core WDs; 
see \citealt{ZS2022} and references therein).

Currently, \markchange{two intact, short-period} giant planets (or planet candidates) 
are known orbiting single WDs. 
The transiting planet WD\,1856+534\,b \citep{Vanderburg+2020} orbits 
a $(0.58 \pm 0.04) \MSol$ WD at a separation of $0.02 \AU$ ($\approx 4 \RSol$). 
Its mass is constrained to be $0.84 \MJ < M_{\rm p} < 13.8 \MJ$ \citep{Vanderburg+2020,Xu+2021}. 
Our {\tt MESA} simulations suggest that a planet would struggle to avoid tidal disruption 
through most of the allowed mass range, 
even if it succeeded in ejecting the envelope with the aid of dust-driven winds. 
For this system, high-eccentricity tidal migration is a plausible alternative formation scenario \citep{MP2020, OLL2021, Stephan+2021}.
Meanwhile, WD\,0141--675 has an astrometric planet candidate 
with mass $9.3^{+2.5}_{-1.1} \MJ$ and semi-major axis $0.17 \AU$ ($\approx 34 \RSol$) 
according to the {\it Gaia}~DR3 astrometric solution \citep{Gaia_Arenou+2022}. 
If confirmed, this system would be difficult to explain as a CE survivor:  
the planet's large orbit implies an insufficient energy budget to disrupt the envelope 
\markchange{\citep[but see][]{BS2011, BS2014, Bear+2021, Chamandy+2021}}. 
\markchange{The microlensing giant planet MOA-2010-BLG-477Lb 
is also associated with a WD \citep{Blackman+2021}. 
Based on its large sky-projected separation from the host ($2.8 \pm 0.5 \AU$), 
this planet likely avoided engulfment during late-stage stellar evolution.}

Over the years, several studies have reported short-period planet candidates 
orbiting horizontal branch stars or hot subdwarfs 
\citep[e.g.][]{Geier+2009, Setiawan+2010, Charpinet+2011}. 
However, none has been confirmed to date \citep[e.g.][]{Norris+2011, JonesJenkins2014, Krzesinski2015}. 
\markchange{Such objects would have survived engulfment during the first RGB
and may have ejected a portion of the stellar envelope 
\citep[e.g.][]{BS2011, BS2014}.} 
We do not find conditions under which this can occur for $M_{\rm p} \leq 10 \MJ$. 
\markchange{Some recent works have suggested that planetary survival is possible 
if engulfment coincides with the host's core helium flash \citep{Bear+2011b, Bear+2021, Merlov+2021}. 
Our {\tt MESA} extension could be used to reexamine this scenario.}

\section{Conclusion} \label{s:Conclusion}

We have studied the evolution of bright giant stars of $1$--$1.5 \MSol$ 
following the engulfment of a Jupiter-sized planet of $1$--$10 \MJ$ using {\tt MESA}. 
We employed 1D spherical approximations of both the heating of the envelope by the planet 
and the resulting hydrodynamical evolution.

For host stars on the first-ascent RGB, as well as for AGB stars with low-mass ($\lesssim 3 \MJ$) planets, 
an engulfed planet causes a mild-to-moderate adjustment of the stellar structure. 
The star brightens by up to $\approx 1$ mag over a few years 
and dims again on a Kelvin--Helmholtz timescale.  
Bright RGB and AGB stars display a prominent `double peak' in the light curve. 
The first peak is associated with hydrogen recombination in the outer layers. 

For late AGB stars, an engulfed planet of sufficient mass 
deposits a major disturbance in the envelope, 
characterized by supersonic expansion of the outer layers.  
In the short term, these systems produce bright, red, dusty eruptions powered by hydrogen recombination, 
similar to LRNe. 
Their long-term evolution is unclear due to a combination of numerical limitations, 
the potential importance of 3D hydrodynamical effects, 
and the uncertain role of dust grains. 
Future works may be able to address some of these issues 
following recent studies of red supergiant stars undergoing pre-supernova outbursts and failed supernovae 
\citep[e.g.][]{RoMatzner2017, Fuller2017, Coughlin+2018, Linial+2021, MatznerRo2021, Tsang+2022}. 
Regardless of the fate of the envelope, 
we find that the planet is always tidally disrupted. 

The optical/infrared transients produced by planetary engulfment events 
may be observed by current and upcoming time-domain facilities. 
The expected Galactic event rates for `weak red transients' and `red novae' 
caused by planetary engulfment are $\sim 0.03 \yr^{-1}$ and $\sim 0.1 \yr^{-1}$, respectively. 
These RNe are dimmer and of longer duration than those caused by stellar mergers. 

We find that short-period, Jupiter-mass planets around WDs are unlikely 
to have arrived in their observed orbits via a CE phase. 
However, we have only considered engulfment events involving a single planet. 
In a multi-planet system, it is possible for the star to engulf several planets successively. 
In principle, our method could be extended to apply to this scenario, 
with each planet acting as an independent heat source embedded in the envelope,
provided the planets' mutual gravitational interactions are negligible. 
Successive engulfment events offer better prospects for envelope ejection: 
even if the first planet cannot eject the envelope, 
it may do enough `damage' that the second ejects the envelope and survives \citep{Chamandy+2021}. 
Multi-planet engulfment events may in fact be common: roughly $50\%$ of cold Jupiters 
have an outer companion of comparable mass \citep{Bryan+2016}. 
A common-envelope origin for massive, short-period giant planets around WDs cannot be categorically dismissed. 

We close by remarking on the main uncertainties and caveats of this work. 
Some obvious limitations arise from the assumption of 1D spherical symmetry 
in both the heating of the envelope by an engulfed planet 
and the dynamical and thermal responses. 
This is an adequate approximation \markchange{while the planet is engulfed well below the stellar surface
and if the stellar response is in the quasistatic regime. 
There may be observable phenomena associated with the grazing phase of an engulfment event that we do not consider.} 
In cases where the envelope is hydrodynamically disrupted,
3D hydrodynamics simulations of the envelope are desirable 
to characterize the evolution of the ejecta \citep[cf.][]{Staff+2016, Tsang+2022}. 
\markchange{We note that several first-ascent RGB stars possess extended, dusty circumstellar disks 
that may have formed during the engulfment of low-mass companions 
\citep[e.g.][]{Jura2003, Zuckerman+2008, Melis+2009, Melis2020}.}

We neglected the transfer of the planet's orbital angular momentum to the stellar envelope. 
This would affect the late stages of inspiral (including prospects for planetary survival) 
and the dynamics of envelope ejection.
We plan to incorporate angular-momentum evolution in a future study using {\tt MESA}.

Finally, we did not consider the fate of planetary debris after tidal disruption. 
\markchange{This may alter the total energy budget of the engulfment process 
(see Section \ref{s:InspPhysics:Energetics}), 
potentially altering the predictions we have made in this work.}
Additional \markchange{mixing and thermal processes} after tidal disruption 
therefore should also be addressed in a future study. 

\begin{acknowledgments}
We thank Jim Fuller, May Gade Pedersen, Logan Prust, and Tin Long Sunny Wong for a number of fruitful discussions, 
as well as the organizers and participants of the KITP program 
{\it White Dwarfs as Probes of the Evolution of Planets, Stars, 
the Milky Way and the Expanding Universe} in Fall 2022. 
We thank the anonymous referee for a thorough, insightful, and speedy review. 

This research was supported in part by the National Science Foundation under grants PHY-1748958 and AST-2107796, 
the Heising-Simons Foundation, and the Simons Foundation (216179, LB). 
CEO gratefully acknowledges a Space Grant Graduate Research Fellowship 
from the New York Space Grant Consortium.
\end{acknowledgments}

\software{{\tt MESA} \citep[r22.05.1;][]{Paxton+2011,Paxton+2013,Paxton+2015,Paxton+2018,Paxton+2019,Jermyn+2022}, {\tt ipython}/{\tt jupyter} \citep{PerezGranger2007,Kluyver+2016}, {\tt NumPy} \citep{Harris+2020}, {\tt matplotlib} \citep{Hunter2007}, {\tt SciPy} \citep{Virtanen+2020}
}

\bibliography{refs}
\bibliographystyle{aasjournals}

\end{document}